\documentclass[twoside,12pt]{article}
\usepackage{epsfig}

\usepackage[a4paper, margin=1.5cm]{geometry}
\usepackage{geometry}                		
\usepackage{graphicx}				
\usepackage{amssymb}
\usepackage{amsmath}
\usepackage{bm}
\usepackage{mathtools}
\usepackage{flexisym}
\usepackage{color}
\usepackage{float}
\usepackage{cancel}
\usepackage{comment}				

\begin{document}

\begin{center}
\LARGE\bf Modification of T/E models and their multi-field versions
\end{center}

\footnotetext{\hspace*{-0cm}\footnotesize MAN Ping Kwan, Ellgan, E-mail:  ellgan101@akane.waseda.jp}

\begin{center}
\rm MAN Ping Kwan, Ellgan$^{\rm a)\dagger}$
\end{center}

\begin{center}
\begin{footnotesize} \sl
$Department \; of \; Pure \; and \; Applied \; Physics, $ \\
$Graduate \; School \; of \; Advanced \; Science \; and \; Engineering, $ \\
$Waseda \; University, \; Tokyo,\; Japan^{\rm a)}$ \\
\end{footnotesize}
\end{center}

\section{Abstract}
\noindent We propose a modification on the super-potential of (arXiv: 1901.09046) which discusses the simplest construction of inflationary $\alpha$ attractor models (also called T/E models) in supergravity without the sinflaton. This can make all the components of the first derivative vanish at the minimum point after the specific choice for the modified part. We also extend T/E models into the multi-field versions, motivated by considering the effect of multi orthogonal nilpotent super-fields, whose sgoldstino, sinflatons and inflatinos are very heavy such that we can take the orthogonal nilpotent constraints to obtain physical limits. Finally, we study the inflation dynamics in the double field cases of both T and E models and show their turning rates and the square mass of the entropic perturbation, which can be used for further observational verification.

\section{Introduction}
\begin{table}[h!]
\begin{center}
\begin{tabular}{ |c|c|c|c| }
\hline
$\text{Slow-roll parameters}$ & $\text{Range(s)}$ & $\text{Spectral indices}$ & $\text{Range(s)}$ \\
\hline
${\epsilon}_{V}$ & $<0.004$ & ${n}_{s} - 1$ & $[-0.0423, -0.0327]$ \\
\hline
${\eta}_{V}$ & $[-0.021, -0.008]$ & $\frac{d {n}_{s}}{d \ln{k}}$ & $[-0.008, 0.012]$ \\
\hline
${\xi}_{V}$ & $[-0.0045, 0.0096]$ & $\frac{d^{2} {n}_{s}}{d \ln{k}^{2}}$ & $[-0.003, 0.023]$ \\
\hline
${H}_{\text{hc}}$ & $< 2.5 \times {10}^{-5} {M}_{\text{pl}}$ & ${V}_{\text{hc}}$ & $< \left( 1.7 \times {10}^{16} \; \text{GeV} \right)^{4}$ \\
\hline
\end{tabular}
\end{center}
\caption{Slow roll potential parameters and spectral indices in Planck 2018}
\label{table:Planck data 2018 slow roll potential parameters and spectral indices}
\end{table}

\noindent Supergravity (SUGRA) has been one of the most promising frameworks to illustrate cosmological inflation because it can be used for deriving Starobinsky model \cite{1307.1137}, which can satisfy the recent Planck observations shortlisted in Table \ref{table:Planck data 2018 slow roll potential parameters and spectral indices} \cite{1807.06211}. Non-linearly realized supersymmetry (SUSY), like Volkov-Akulov (VA) SUSY \cite{Volkov_Akulov}, carries much weight in model construction with phenomenological de Sitter (dS) vacua. Global VA SUSY is based on a nilpotent super-field condition in \cite{Rocek}, while other constrained super-fields are well-studied in \cite{0907.2441}. It was later discovered that these constrained properties can be extended to local SUSY \cite{1511.07547, 1507.08619}, and they are advantageous to dS spacetime construction \cite{1507.08264, 1509.02136, 1509.02137} and inflation model construction \cite{1403.3269, 1408.4096, 1411.2605}. \cite{1509.06345, 1603.02653, 1603.02661} discuss the SUGRA construction of orthogonal nilpotent super-fields and show that these constraints are the results of having a finite limit when the mass scales of the super-fields involved are very large, while \cite{1512.00545} discusses the application of orthogonal nilpotent super-fields. Furthermore, \cite{1411.1121, 1502.07627, 1705.09247, 1707.08678} show that a nilpotent property in SUGRA can be embedded in a $\overline{D3}$ brane in superstring theory to give another origin of the production of uplifting in KKLT model \cite{hep-th/0301240}. 

\vspace{3mm}

\noindent \cite{1603.02661, 1512.00545} show that both sgoldstino and sinflaton can be expressed by the bilinear spinor forms of a goldstino when super-fields $\bold{S}: \; \left\{ {S}_{R} + i {S}_{I}, {P}_{L} {\Omega}^{S}, {F}^{S} \right\}$ and $\bold{\Phi}: \; \left\{ {\phi}_{R} + i {\phi}_{I}, {P}_{L} {\Omega}^{\phi}, {F}^{\phi} \right\}$ are subject to constraints
\begin{equation}
\bold{S}^{2} = 0, \quad \bold{S} \bold{B} = 0, \\
\end{equation}
\noindent where $\bold{B} = \frac{1}{2 i} \left( \bold{\Phi} - \overline{\bold{\Phi}} \right)$ and the auxiliary field ${F}^{S}$ is not vanishing, as
\begin{equation}
S := {S}_{R} + {S}_{I} = \frac{\overline{{\Omega}^{S}} {P}_{L} {\Omega}^{S} }{2 {F}^{S} }, \quad \quad {F}^{S} \neq 0, \\
\end{equation}
\noindent and ${\phi}_{I}$ is given by \cite{1603.02661, 1901.09046}
\begin{equation}
\begin{split}
{\phi}_{I} =&\; \frac{i}{4} \left\{ \left( \frac{ \overline{{\Omega}^{S}} }{ \overline{{F}^{S}} } \right) {\gamma}^{\mu} {P}_{L} \left( \frac{ {\Omega}^{S} }{ {F}^{S} } \right) - \frac{\overline{S} }{\overline{{F}^{S}} } \left( {\mathcal{D}}_{\nu} \frac{ \overline{{\Omega}^{S}} }{ {F}^{S} } \right) {\gamma}^{\mu} {\gamma}^{\nu} {P}_{L} \frac{{\Omega}^{S} }{ {F}^{S} } \right. \\
&\left. - \frac{\left| S \right|^{2} }{2 \left| {F}^{S} \right|^{2} } \left( \mathcal{D}_{\nu} \frac{ \overline{{\Omega}^{S}} }{ \overline{{F}^{S}} } \right) \left( {\gamma}^{{\mu}{\nu}{\rho}} + {\gamma}^{\nu} {\eta}^{{\nu}{\rho}} \right) {P}_{L} \left( {\mathcal{D}}_{\rho} \frac{ {\Omega}^{S} }{ {F}^{S} } \right) - \text{c.c.} \right\} {\mathcal{D}}_{\mu} {\phi}_{R}. \\
\end{split}
\end{equation}

\noindent where c.c. means complex conjugate. The origin of $\bold{S}^{2} = 0$ and $\bold{SB} = 0$ comes from the fact that mass scales of $S$ and ${\phi}_{I}$ are very heavy \cite{1603.02661} such that $S$ and ${\phi}_{I}$ do not vary during the inflation evolution. To realize this, we consider the correction on the K\"ahler potential as \cite{1603.02661}
\begin{equation}
\frac{\Delta K}{ {M}_{\text{pl}}^{2} } = - {c} \left( S \overline{S} \right)^{2} - \frac{1}{4} \left( {c}_{1} S + \overline{{c}_{1}} \overline{S} \right) \left( \Phi - \overline{\Phi} \right)^{2} + \frac{1}{4} {c}_{2} S \overline{S} \left( \Phi - \overline{\Phi} \right)^{2}, \\
\end{equation}

\noindent where $c, \; {c}_{2} \in \mathbb{R}$ and ${c}_{1} \in \mathbb{C}$, $S := {S}_{R} + i {S}_{I}$ and $\Phi := {\phi}_{R} + i {\phi}_{I}$ are the lowest scalar components of the super-fields $\bold{S}$ and $\bold{\Phi}$, and\footnote{In this paper, the lowest scalar component of super-field $\bold{\Phi}$ is $\Phi = Z = {Z}_{R} + i {Z}_{I} = \text{Re}\left( Z \right) + i \text{Im}\left( Z \right)$ for T model and $\Phi = T = {T}_{R} + i {T}_{I} = \text{Re}\left( T \right) + i \text{Im}\left( T \right)$ for E model. } these super-fields have not yet been subject to the constraints\footnote{In this subsection, we do not treat the lowest scalar components of $\bold{S}$ and $\bold{\Phi}$ as fermion bilinear forms. } $\bold{S}^{2} = 0$ and $\bold{SB} = 0$. Thus, the total K\"ahler potential $K$ and total super-potential $W$ of T model can be considered as

\begin{equation}
K = {K}_{\text{T model}} + \Delta K, \quad \quad W = {W}_{0} {e}^{S}, \\
\end{equation}

\noindent where ${K}_{\text{T model}}$ is given by Eq.(\ref{Kahler potential of T model})\footnote{\label{note} In order not to distract readers from reading and understanding, for T and E models from orthogonal constrained super-fields, please refer to Appendix \ref{A review: Single orthogonal constrained field}. }. The square mass matrix\footnote{The square mass of each field is given by $\left( {M}^{2} \right)^{I}_{\;K} := {G}^{IJ} {\nabla}_{J} {\nabla}_{K} {V}_{F} = {G}^{IJ} \left( {\partial}_{J} {\partial}_{K} - {\Gamma}^{L}_{JK} {\partial}_{L} \right) {V}_{F} $, where ${G}_{IJ}$ is the metric of the field space. } of the $F$ term potential evaluated at $S = 0$ and $Z = 0$ is diagonal with the following elements in the ordered basis $\left\{ \text{Re} \left( Z \right), \text{Im} \left( Z \right), \text{Re} \left( S \right), \text{Im} \left( S \right) \right\}$
\begin{equation}
\begin{split}
&\frac{2 f'(0)}{3 \alpha {M}_{\text{pl}}^{4} }, \quad \frac{1}{{M}_{\text{pl}}^{4}} \left\{ \frac{2 \left| {c}_{1} \right|^{2} \left| {F}_{S} \right|^{4} }{9 \alpha^2 \left| {W}_{0} \right|^{2} }-\frac{2 \overline{{c}_{1}} \left| {F}_{S} \right|^{2} }{3 \alpha }+\frac{2 {c}_{2} \left| {F}_{S} \right|^{4} }{3 \alpha  \left| {W}_{0} \right|^{2}} - \frac{2 {c}_{1}
  \left| {F}_{S} \right|^{2} }{3 \alpha } + 4 \left| {F}_{S} \right|^{2} - 4 \left| {W}_{0} \right|^{2}+\frac{2 f'(0)}{3 \alpha } \right\}, \\
 &\frac{1}{{M}_{\text{pl}}^{4}} \left\{ \frac{8 c \left| {F}_{S} \right|^{6} }{ \left| {W}_{0} \right|^{4} }+\frac{4 \left| {F}_{S} \right|^{4} }{\left| {W}_{0} \right|^{2}}-2 \left| {F}_{S} \right|^{2} - 4 \left| {W}_{0} \right|^{2} \right\}, \quad \frac{1}{{M}_{\text{pl}}^{4}} \left\{ \frac{8 c \left| {F}_{S} \right|^{6} }{ \left| {W}_{0} \right|^{4} }+2 \left| {F}_{S} \right|^{2}-4 \left| {W}_{0} \right|^{2} \right\}. \\
   \end{split}
\end{equation}

\noindent Apart from this, the total K\"ahler potential $K$ and total super-potential $W$ of E model can be considered as
\begin{equation}
K = {K}_{\text{E model}} + \Delta K, \quad \quad W = {W}_{0} {e}^{S}. \\
\end{equation}

\noindent where ${K}_{\text{E model}}$ is given by Eq.(\ref{Kahler potential of E model}). The square mass matrix of the $F$ term potential at $S = 0$ and $T = 1$ is diagonal with the following elements in the ordered basis $\left\{ \text{Re} \left( T \right), \text{Im} \left( T \right), \text{Re} \left( S \right), \text{Im} \left( S \right) \right\}$
\begin{equation}
\begin{split}
&\frac{4 h''(0)}{3 \alpha {M}_{\text{pl}}^{4} }, \quad \frac{1}{ {M}_{\text{pl}}^{4} } \left\{ \frac{32 \left| {c}_{1} \right|^{2} \left| {F}_{S} \right|^{4} }{9 \alpha^2 \left| {W}_{0} \right|^{2}} - \frac{8 \overline{{c}_{1}} \left| {F}_{S} \right|^{2} }{3 \alpha }+\frac{8 {c}_{2} \left| {F}_{S} \right|^{4} }{3 \alpha  \left| {W}_{0} \right|^{2}}-\frac{8 {c}_{1} \left| {F}_{S} \right|^{2} }{3 \alpha }+4 \left| {F}_{S} \right|^{2} -4 \left| {W}_{0} \right|^{2} \right\}, \\
&\frac{1}{ {M}_{\text{pl}}^{4} } \left\{ \frac{8 c \left| {F}_{S} \right|^{6} }{ \left| {W}_{0} \right|^{4} }+\frac{4 \left| {F}_{S} \right|^{4} }{\left| {W}_{0} \right|^{2}} - 2 \left| {F}_{S} \right|^{2} - 4 \left| {W}_{0} \right|^{2} \right\}, \quad \frac{1}{ {M}_{\text{pl}}^{4} } \left\{ \frac{8 c \left| {F}_{S} \right|^{6} }{\left| {W}_{0} \right|^{4} }+2 \left| {F}_{S} \right|^{2}-4 \left| {W}_{0} \right|^{2} \right\}. \\
\end{split}
\end{equation}
\noindent given that $h \left( 0 \right) = h' \left( 0 \right) = 0$. Hence, one can see that when ${c}, \; {c}_{1}, \; {c}_{2} \rightarrow \infty$, the masses of $\text{Im}\left( T \right), \text{Re}\left( S \right), \text{Im}\left( S \right)$ for T model, and that of $\text{Im}\left( T \right), \text{Re}\left( S \right), \text{Im}\left( S \right)$ for E model are very large such that they are stabilized at their corresponding minimum points. \cite{1603.02661} shows the derivation of orthogonal nilpotent constraints $\bold{S}^{2} = 0$ and $\bold{SB} = 0$ by sending the masses of the sgoldstino, inflatino and sinflaton to infinity, which physically means that the stabilization of the sgoldstino, inflatino and sinflaton leads to orthogonal nilpotent constraints. Since they are much larger than the Hubble scale as ${c}, \; {c}_{1}, \; {c}_{2} \rightarrow \infty$, we can assume they are stabilized throughout the inflation process and ${Z}_{R}$ (${T}_{R}$) governs the inflation dynamics in T model (E model). \cite{1901.09046} gives the simple construction of T and E models and verifies the predictions by the observation data of the ${n}_{s} - r$ graph.

\subsection{The problem}
\noindent The $F$ term potential of T model evaluated at the minimum point is
\begin{equation}
\left. {V}_{F} \right|_{0} = \Lambda + \frac{f \left( 0 \right)}{{M}_{\text{pl}}^{2}}, \\
\end{equation}
\noindent where $\Lambda = {M}_{\text{pl}}^{-2} \left[ \left| {F}_{S} \right|^{2} - 3 \left| {W}_{0} \right|^{2} \right]$ is a cosmological constant \cite{1901.09046}. Since the first derivatives of T model evaluated at the minimum point are
\begin{equation}
{M}_{\text{pl}}^{-2} \left\{ 0, 0, 2 \left( {M}_{\text{pl}}^{2} {\Lambda} + \left| {W}_{0} \right|^{2} + f \left( 0 \right) \right), 0 \right\}, \\
\end{equation}

\noindent in the ordered basis $\left\{ \text{Re} \left( Z \right), \text{Im}\left( Z \right), \text{Re} \left( S \right), \text{Im} \left( S \right) \right\}$. Since $S$ has not been taken as a bilinear spinor form, as a scalar, $S$ should be stabilized at the minimum point. To satisfy $\left. {V}_{F} \right|_{0} = \Lambda$,  $f \left( 0 \right)$ is taken as zero, which implies ${\Lambda} = - \left| {W}_{0} \right|^{2} / {M}_{\text{pl}}^{2} \leq 0$ as a contradiction to the assumption that the spacetime is dS during inflation. In fact, this situation also appears in E model. The $F$ term potential of E model evaluated at the minimum point is
\begin{equation}
\left. {V}_{F} \right|_{0} = \Lambda + \frac{h \left( 0 \right)}{{M}_{\text{pl}}^{2}}, \\
\end{equation}
\noindent and its first derivative evaluated at the minimum point are
\begin{equation}
{M}_{\text{pl}}^{-2} \left\{ - h' \left( 0 \right), 0, 2 \left( {M}_{\text{pl}}^{2} {\Lambda} + \left| {W}_{0} \right|^{2} + h \left( 0 \right) \right), 0 \right\}. \\
\end{equation}

\noindent in the ordered basis $\left\{ \text{Re} \left( T \right), \text{Im}\left( T \right), \text{Re} \left( S \right), \text{Im} \left( S \right) \right\}$, where $h \left(0 \right) = h' \left(0 \right) = 0$. Thus, the same situation for $\Lambda$ still comes out. Hence, we propose a modified solution to solve this problem. 

\vspace{3mm}

\noindent The organization of this paper is the following. In section \ref{A solution to non-vanishing first derivative of VF}, we propose a solution to solve the non-vanishing problem of the first derivative by adding a constant term in the super-potential. This also enables us to tune the value of the constant so as to obtain the de-Sitter space-time at the minimum point, which is phenomenologically essential for the end of inflation. In section \ref{Constructions of potential by multi orthogonal constrained fields}, we extend the modified construction to find the multi-field versions of T/E models by considering there are $n$ multi orthogonal super-fields instead of one, while the number of nilpotent super-fields remains one since it is associated with a goldstino in 4 dimensional $\mathcal{N}=1$ SUGRA. In section \ref{Properties of minimum point(s)}, we study the properties of the minimum point in multi-field versions. In section \ref{The basic setup of multi constrained fields} and \ref{Numerical calculations}, we show the double field inflation dynamics of both T and E models and evaluate the scales of turn rate (per Hubble parameter) and square mass of the entropic perturbation (or called iso-curvature mode). We discuss our results in section \ref{Discussion} and finally conclude in section \ref{Conclusions}.

\subsection{A solution to non-vanishing first derivative of ${V}_{F}$}
\label{A solution to non-vanishing first derivative of VF}
\noindent The K\"ahler potentials are Eq.(\ref{Kahler potential of T model}) for T model and Eq.(\ref{Kahler potential of E model}) for E model, while the super-potential of both T and E models is given by\footnote{For details of derivation, please refer to Appendix \ref{A derivation of the modified constant}. }
\begin{equation}
\label{solution of modified super-potential}
W = {W}_{0} e^{S} - \left( \delta + \frac{1}{3} \right) {W}_{0}. \\
\end{equation}

\noindent In this case, the $F$ term potential of T model at $S = {Z}_{I} = 0$ becomes
\begin{equation}
{V}_{\text{T model}} = \frac{ \left| {F}_{S} \right|^{2} - 3 \left| {W}_{0} + k \right|^{2} }{ {M}_{\text{pl}}^{2} } + \frac{ f \left( {Z}_{R}^{2} \right)}{{M}_{\text{pl}}^{2}} = \delta \left( 2 - 3 \delta \right) \frac{\left| {W}_{0} \right|^{2}}{{M}_{\text{pl}}^{2}} + \frac{f \left( {Z}_{R}^{2} \right)}{{M}_{\text{pl}}^{2}}, \\
\end{equation}

\noindent This modification can allow a small cosmological constant $\Lambda$ with arbitrary scales of $\left| {W}_{0} \right|^{2}$ (and $\left| {F}_{S} \right|^{2}$). We find that the (analytical) minimum of T model is $\left( {Z}_{R}, {Z}_{I}, {S}_{R}, {S}_{I} \right) = \left( 0,0,0,0 \right)$ and that of E model is $\left( {T}_{R}, {T}_{I}, {S}_{R}, {S}_{I} \right) = \left( 1,0,0,0 \right)$\footnote{One can expand the $F$ term potential in terms of general field coordinates and obtain the derivatives to confirm whether there are other optimum field coordinates. The fine tuned optimum value can appear if we finely tune the parameters inside. However, since $c$, ${c}_{1}$ and ${c}_{2}$ are finite, large and underdetermined, even if we can find a set of parameters given $c$, ${c}_{1}$ and ${c}_{2}$, those parameters will no longer provide the same minimum point when either $c$, ${c}_{1}$ or ${c}_{2}$ is changed. Thus, for phenomenological discussion, we consider the minimum point of T model is $\left( {Z}_{R}, {Z}_{I}, {S}_{R}, {S}_{I} \right) = \left( 0,0,0,0 \right)$ while $\left( {T}_{R}, {T}_{I}, {S}_{R}, {S}_{I} \right) = \left( 1,0,0,0 \right)$ for E model so that no matter how large we take $c$, ${c}_{1}$ and ${c}_{2}$, the minimum point is unchanged. }. The cosmological constant and $\left| {F}_{S} \right|^{2}$ are given by

\begin{equation}
\label{A result of a cosmological constant condition}
\begin{split}
\Lambda =&\; - \frac{\overline{W}_{0} }{ {W}_{0} } \frac{\left( {W}_{0} + k \right) \left( {W}_{0} + 3 k \right)}{{M}_{\text{pl}}^{2}} = \delta \left( 2 - 3 \delta \right) \frac{\left| {W}_{0} \right|^{2}}{{M}_{\text{pl}}^{2}}, \\
\left| {F}_{S} \right|^{2} =&\; 2 \left( k + {W}_{0} \right) \overline{{W}_{0}} = \frac{2}{3} \left( 2 - 3 \delta \right) \left| {W}_{0} \right|^{2}. \\
\end{split}
\end{equation}

\noindent All components of the first derivatives evaluated at the minimum point become zero. The mass matrix evaluated at the minimum point becomes diagonal and the diagonal elements are
\begin{equation}
\begin{split}
&\frac{2 f'(0)}{ 3 \alpha {M}_{\text{pl}}^{4} }, \quad \frac{2 f'(0)}{ 3 \alpha {M}_{\text{pl}}^{4} } + \frac{4 (2 - 3 \delta) \left| {W}_{0} \right|^{2} }{81 \alpha^2 {M}_{\text{pl}}^{4} } \left[ {c}_{1I}^{2} ( 4 - 6 \delta ) + {c}_{1R}^{2} (4-6 \delta )+6 \alpha  {c}_{1R} (3 \delta - 2) \right. \\
&\left. + 3 \alpha  (9 \alpha  \delta +12 \alpha - 6 {c}_{2} \delta +4 {c}_{2}) \right], \quad \frac{8}{27} (2 - 3 \delta )^2 \frac{\left| {W}_{0} \right|^{2}}{{M}_{\text{pl}}^{4}} \left[ 8 c (2- 3 \delta ) + 3 \right], \quad \frac{64}{27} c (2 - 3 \delta)^3 \frac{\left| {W}_{0} \right|^{2}}{{M}_{\text{pl}}^{4}}. \\
\end{split}
\end{equation}

\noindent where ${c}_{1} = {c}_{1R} + i {c}_{1I}$. It is trivial to find that the third and fourth diagonal values are positive if $c>0$ and $0 < \delta < \frac{2}{3}$. The second diagonal value is equal to
\begin{equation}
\frac{4 \left| {W}_{0} \right|^{2} }{ 81 \alpha^2 {M}_{\text{pl}}^{4} } \left( 2 - 3 \delta \right) \left\{ 2 {c}_{1I}^{2} \left( 2 - 3 \delta \right) + 2 {c}_{2} \left( 2 - 3 \delta \right) + 2 \left( 2 - 3 \delta \right) \left( {c}_{1R} - \frac{3}{2} \alpha \right)^{2} + \frac{27 {\alpha}^{2} }{2} \left( 3 \delta + 2 \right) \right\}, \\
\end{equation}

\noindent which is also positive. $f ' \left( 0 \right) \geq 0$ so as to obtain the dS spacetime. Apart from this, the $F$ term potential of E model at $S = {T}_{I} = 0$ becomes

\begin{equation}
{V}_{\text{E model}} = \frac{\left| {F}_{S} \right|^{2} - 3 \left| {W}_{0} + k \right|^{2}}{{M}_{\text{pl}}^{2}} + \frac{h \left( 1 - {T}_{R} \right)}{ {M}_{\text{pl}}^{2} } = \delta \left( 2 - 3 \delta \right) \frac{\left| {W}_{0} \right|^{2}}{{M}_{\text{pl}}^{2}} + \frac{h \left( 1 - {T}_{R} \right)}{{M}_{\text{pl}}^{2}}. \\
\end{equation}

\noindent All components of the first derivatives become zero and the mass matrix evaluated at the minimum point becomes diagonal, whose diagonal elements are
\begin{equation}
\begin{split}
&\frac{4 h''(0)}{3 \alpha {M}_{\text{pl}}^{4}}, \quad \frac{4 (2 - 3 \delta) \left| {W}_{0} \right|^{2} }{81 \alpha^2 {M}_{\text{pl}}^{4} } \left\{ {c}_{1I}^2 (64-96 \delta ) + {c}_{1R}^2 (64 - 96 \delta )+24 \alpha {c}_{1R} (3 \delta -2) + 3 \alpha  \left[ 3 \alpha  (3 \delta + 4) \right. \right. \\
&\left. \left. + 8 {c}_{2} (2-3 \delta ) \right] \right\}, \quad \frac{8}{27} (2 - 3 \delta )^2 \frac{\left| {W}_{0} \right|^{2}}{ {M}_{\text{pl}}^{4} } \left[ 8 c (2 - 3 \delta ) + 3 \right], \quad \frac{64}{27} c (2 - 3 \delta)^3 \frac{\left| {W}_{0} \right|^{2}}{ {M}_{\text{pl}}^{4} }. \\
\end{split}
\end{equation}

\noindent where ${c}_{1} = {c}_{1R} + i {c}_{1I}$. It is trivial to find that the third and fourth diagonal values are positive if $c>0$ and $0 < \delta < \frac{2}{3}$. The second diagonal value is equal to
\begin{equation}
\frac{4 \left| {W}_{0} \right|^{2} }{ 81 \alpha^2 {M}_{\text{pl}}^{4} } \left( 2 - 3 \delta \right) \left\{ 32 {c}_{1I}^{2} \left( 2 - 3 \delta \right) + 24 \alpha {c}_{2} \left( 2 - 3 \delta \right) + 32 \left( 2 - 3 \delta \right) \left( {c}_{1R} - \frac{3}{8} \alpha \right)^{2} + \frac{27 {\alpha}^{2} }{2} \left( 3 \delta + 2 \right) \right\}, \\
\end{equation}

\noindent which is also positive. $h '' \left( 0 \right) \geq 0$ so as to obtain the dS spacetime. Next, we are going to investigate the multi orthogonal constrained fields of the above modified models.

\section{Constructions of potential by multi orthogonal constrained fields}
\label{Constructions of potential by multi orthogonal constrained fields}
\noindent We construct the potential of multi-inflaton multiplets where their sinflatons can be expressed by the fermions of the super-fields. Given that $\bold{S}: \; \left\{ {S}_{R} + i {S}_{I}, {P}_{L} {\Omega}^{S}, {F}^{S} \right\}$ and $\bold{\Phi}_{l}: \; \left\{ {\phi}_{lR} + i {\phi}_{lI}, {P}_{L} {\Omega}^{{\phi}_{l}}, {F}^{{\phi}_{l}} \right\}$ by
\begin{equation}
\bold{S}^{2} = 0, \quad \quad \bold{S} \bold{B}_{l} = 0, \\
\end{equation}
\noindent where $\bold{B}_{l} = \frac{1}{2 i} \left( \bold{\Phi}_{l} - \overline{\bold{\Phi}_{l}} \right)$ $\forall l \in \left\{1,\cdots, n \right\}$, on solving by the same technique, we have
\begin{equation}
S := {S}_{R} + i {S}_{I} = \frac{ \overline{{\Omega}^{S}} {P}_{L} {\Omega}^{S} }{ 2 {F}^{S} }, \quad \quad {F}^{S} \neq 0, \\
\end{equation}
\noindent and ${\phi}_{lI}$ are given by
\begin{equation}
\begin{split}
{\phi}_{lI} =&\; \frac{i}{4} \left\{ \left( \frac{ \overline{{\Omega}^{S}} }{ \overline{{F}^{S}} } \right) {\gamma}^{\mu} {P}_{L} \left( \frac{ {\Omega}^{S} }{ {F}^{S} } \right) - \frac{\overline{S} }{\overline{{F}^{S}} } \left( {\mathcal{D}}_{\nu} \frac{ \overline{{\Omega}^{S}} }{ {F}^{S} } \right) {\gamma}^{\mu} {\gamma}^{\nu} {P}_{L} \frac{{\Omega}^{S} }{ {F}^{S} } \right. \\
&\left. - \frac{\left| S \right|^{2} }{2 \left| {F}^{S} \right|^{2} } \left( \mathcal{D}_{\nu} \frac{ \overline{{\Omega}^{S}} }{ \overline{{F}^{S}} } \right) \left( {\gamma}^{{\mu}{\nu}{\rho}} + {\gamma}^{\nu} {\eta}^{{\nu}{\rho}} \right) {P}_{L} \left( {\mathcal{D}}_{\rho} \frac{ {\Omega}^{S} }{ {F}^{S} } \right) - \text{c.c.} \right\} {\mathcal{D}}_{\mu} {\phi}_{lR}. \\
\end{split}
\end{equation}
\noindent where c.c. means complex conjugate.

\subsection{The origin of the constrained super-fields in the multi-field case}
\subsubsection{T model}
\noindent Similar to the above, we are going to realize the physical origin of $\bold{S}^{2} = 0$ and $\bold{S} \bold{B}_{l} = 0$ in the multi-field case. We consider the correction on the K\"ahler potential as
\begin{equation}
\label{Kahler correction of the multi-field case}
\frac{\Delta K}{ {M}_{\text{pl}}^{2} } = - c \left( S \overline{S} \right)^{2} - \frac{1}{4} \sum_{l=1}^{n} \left( {c}_{1l} S + \overline{{c}_{1l}} \overline{S} \right) \left( {\Phi}_{l} - \overline{{\Phi}_{l}} \right)^{2} + \frac{1}{4} S \overline{S} \sum_{l=1}^{n} {c}_{2l} \left( {\Phi}_{l} - \overline{{\Phi}_{l}} \right)^{2}, \\
\end{equation}
\noindent where $c, {c}_{2l} \in \mathbb{R}$, ${c}_{1l} \in \mathbb{C}$ $\forall \; l \in \left\{ 1, \cdots, n \right\}$ and ${\Phi}_{l} = {Z}_{l}$ for T model and ${\Phi}_{l} = {T}_{l}$ for E model. The total K\"ahler potential $K$ and total super-potential $W$ of T model become
\begin{equation}
K = {K}_{\text{T model}} + \Delta K, \quad \quad W = {W}_{0} {e}^{S} - \left( \delta + \frac{1}{3} \right) {W}_{0}, \\
\end{equation}

\noindent and the square mass matrix of the $F$ term potential evaluated at $S=0$ and ${Z}_{l} = 0$ is diagonal with the following elements in the ordered basis $\left\{ {Z}_{lR}, {Z}_{lI}, {S}_{R}, {S}_{I} \right\}$, after the substitution of the first of Eq.(\ref{A cosmological constant condition}) and the second of Eq.(\ref{A result of a cosmological constant condition})
\begin{equation}
\begin{split}
&\left. \frac{2}{3 {\alpha}_{l} {M}_{\text{pl}}^{4} } \frac{d f \left( {Z}_{1} \overline{{Z}_{1}}, \cdots, {Z}_{n} \overline{{Z}_{n}} \right) }{d \left( {Z}_{l} \overline{{Z}_{l}} \right)} \right|_{0}, \left. \frac{2}{3 {\alpha}_{l} {M}_{\text{pl}}^{4} } \frac{d f \left( {Z}_{1} \overline{{Z}_{1}}, \cdots, {Z}_{n} \overline{{Z}_{n}} \right) }{d \left( {Z}_{l} \overline{{Z}_{l}} \right)} \right|_{0} + \frac{4 (2 - 3 \delta) \left| {W}_{0} \right|^{2} }{81 {\alpha}_{l}^2 {M}_{\text{pl}}^{4} } \left\{ {c}_{1lI}^2 (4-6 \delta ) \right. \\
&\left. + {c}_{1lR}^2 ( 4 - 6 \delta )+6 {\alpha}_{l} {c}_{1lR} (3 \delta -2)+3 {\alpha}_{l} (9 {\alpha}_{l} \delta +12 {\alpha}_{l}-6 \text{c23} \delta +4 {c}_{2l}) \right\}, \\
   & \frac{8}{27} \frac{\left| {W}_{0} \right|^{2}}{{M}_{\text{pl}}^{4}} \left( 2 - 3 \delta \right)^{2} \left[ 8 c ( 2 - 3 \delta) + 3 \right], \quad \frac{64}{27} \frac{\left| {W}_{0} \right|^{2}}{{M}_{\text{pl}}^{4}} c \left( 2 - 3 \delta \right)^{3}. \\
   \end{split}
\end{equation}

\noindent where ${c}_{1l} = {c}_{1lR} + i {c}_{1lI}$ and $\left. \frac{d f \left( {y}_{1}, \cdots, {y}_{n} \right)}{d {y}_{l} } \right|_{0} \geq 0$. Hence, one can see that $c$ is responsible for the mass scale of $S$, while $\forall l \in \left\{1,\cdots, n \right\}$, ${c}_{1l}, \overline{{c}_{1l}}, {c}_{2l}$ are responsible for that of ${Z}_{lI}$. If $c, {c}_{1l}, {c}_{2l} \rightarrow \infty$, the mass scales of ${Z}_{lI}, {S}_{R}, {S}_{I}$ evaluated at the minimum point will be very large, so that we can assume that they are fixed and we can take the constraints of $\bold{S}$ and $\bold{B}_{l} = \frac{1}{2i} \left( {\Phi}_{l} - \overline{{\Phi}_{l}} \right)$ to obtain a finite limit as mentioned in \cite{1603.02661}.

\subsubsection{E model}
\noindent For E model, we can consider the same correction as Eq.(\ref{Kahler correction of the multi-field case}). The total K\"ahler potential $K$ and the total super-potential $W$ are given by
\begin{equation}
K = {K}_{\text{E model}} + \Delta K, \quad \quad W = {W}_{0} {e}^{S} - \left( \delta + \frac{1}{3} \right) {W}_{0}. \\
\end{equation}
\noindent The mass matrix of the $F$ term potential evaluated at $S=0$ and ${T}_{l} = 1$ is diagonal with the following elements in the ordered basis $\left\{ {T}_{lR}, {T}_{lI}, {S}_{R}, {S}_{I} \right\}$ after the substitution of the first of Eq.(\ref{A cosmological constant condition}) and the second of Eq.(\ref{A result of a cosmological constant condition})

\begin{equation}
\begin{split}
&\left. \frac{4}{3 {\alpha}_{l} {M}_{\text{pl}}^{4} } \frac{d^2 h \left( {y}_{1}, \cdots, {y}_{n} \right)}{d {y}_{l}^2} \right|_{0}, \quad \frac{4 (3 \delta - 2) \left| {W}_{0} \right|^{2} }{81 {\alpha}_{l}^2 {M}_{\text{pl}}^{4} } \left\{ 32 {c}_{1lI}^2 (3 \delta - 2 ) + 32 {c}_{1lR}^2 (3 \delta - 2 ) + 24 {\alpha}_{l} {c}_{1lR} ( 2 - 3 \delta ) \right. \\
   &\left. -3 {\alpha}_{l} (9 {\alpha}_{l} \delta +12 {\alpha}_{l}-24 {c}_{2l} \delta +16 {c}_{2l}) \right\}, \quad \frac{8}{27} \frac{\left| {W}_{0} \right|^{2}}{ {M}_{\text{pl}}^{4} } (2 - 3 \delta )^2 \left[ 8 c ( 2 - 3 \delta) + 3 \right] , \quad \frac{64}{27} \frac{\left| {W}_{0} \right|^{2}}{ {M}_{\text{pl}}^{4} } c ( 2 - 3 \delta)^3. \\
   \end{split}
\end{equation}

\noindent where ${c}_{1l} = {c}_{1lR} + i {c}_{1lI}$. Hence, one can see that $c$ is responsible for the mass scale of $S$, while $\forall l \in \left\{1,\cdots, n \right\}$, ${c}_{1l}, \overline{{c}_{1l}}, {c}_{2l}$ are responsible for that of ${T}_{lI}$. If $c, {c}_{1l}, {c}_{2l} \rightarrow \infty$, the mass scales of ${T}_{lI}, {S}_{R}, {S}_{I}$ evaluated at the minimum point will be very large, so that we can assume that they are fixed and we can take the constraints of $\bold{S}$ and $\bold{B}_{l} = \frac{1}{2i} \left( {\Phi}_{l} - \overline{{\Phi}_{l}} \right)$ to obtain a finite limit as mentioned in \cite{1603.02661}.

\subsection{A construction of multi-field T model}
\noindent The $F$ term potential of T model with K\"ahler potential $K$ and super-potential $W$ is given by
\begin{equation}
\frac{ {K}_{\text{T model}} }{ {M}_{\text{pl}}^{2} } = - \frac{1}{2} \sum_{l=1}^{n} \frac{ 3 {\alpha}_{l} \left( {Z}_{l} - \overline{{Z}_{l}} \right)^{2} }{ \left( 1 - {Z}_{l} \overline{{Z}_{l}} \right)^{2} } + \frac{\left| {W}_{0} \right|^{2} }{ \left| {F}_{S} \right|^{2} + f \left( {Z}_{1} \overline{{Z}_{1}}, \cdots, {Z}_{n} \overline{{Z}_{n}} \right) } S \overline{S}, \quad \quad W = {W}_{0} {e}^{S} - \left( \delta + \frac{1}{3} \right) {W}_{0}, \\
\end{equation}

\noindent where $f \left( {Z}_{1} \overline{{Z}_{1}}, \cdots, {Z}_{n} \overline{{Z}_{n}} \right)$ is a real function with $f\left( 0, \cdots, 0 \right) = 0$, becomes
\begin{equation}
{V}_{\text{T model}} = \Lambda + \frac{f \left( {Z}_{1}^{2}, \cdots, {Z}_{n}^{2} \right)}{ {M}_{\text{pl}}^{2} }. \\
\end{equation}

\noindent where $\Lambda = \delta \left( 2 - 3 \delta \right) \frac{ \left| {W}_{0} \right|^{2} }{ {M}_{\text{pl}}^{2} }$. When we take\footnote{We have tried to take $f \left( {Z}_{1} \overline{{Z}_{1}}, \cdots, {Z}_{n} \overline{{Z}_{n}} \right) = \prod_{l=1}^{n} {g}_{l} \left( {Z}_{l} \overline{{Z}_{l}} \right) = A \prod_{l=1}^{n} \left( {Z}_{l} \overline{{Z}_{l}} \right)^{2 {n}_{l} }$, where $A$ is a real constant characterizing the potential energy scale. But, it cannot be restored to the single field case Eq.(\ref{Single field T model}) after substituting $\text{Re} \left( {Z}_{l} \right) = \tanh{\left( \frac{ {\phi}_{l} }{ {M}_{\text{pl}} \sqrt{6 {\alpha}_{l} } } \right)}$ and taking the minimum coordinates ${\phi}_{l} = 0$ for all $l>1$. } 
\begin{equation}
\label{A special function of T model}
f \left( {Z}_{1} \overline{{Z}_{1}}, \cdots, {Z}_{n} \overline{{Z}_{n}} \right) = \sum_{l=1}^{n} {g}_{l} \left( {Z}_{l} \overline{{Z}_{l}} \right) = \sum_{l=1}^{n} {m}_{l}^{2} \left( {Z}_{l} \overline{{Z}_{l}} \right)^{2 {n}_{l} }, \\
\end{equation}
\noindent in terms of canonical fields ${\phi}_{l}$, where $\text{Re}{\left( {Z}_{l} \right)} = \tanh{\left( \frac{ {\phi}_{l} }{ {M}_{\text{pl}} \sqrt{6 {\alpha}_{l} }} \right)}$, and take ${Z}_{lI} = 0$, it gives the multi-field T model
\begin{equation}
\label{A Multi-field T model}
{V}_{\text{T model}} = \Lambda + {M}_{\text{pl}}^{-2} \sum_{l=1}^{n} {m}_{l}^{2} \tanh^{2{n}_{l}}{\left( \frac{ {\phi}_{l} }{ {M}_{\text{pl}} \sqrt{6 {\alpha}_{l} } } \right)}. \\
\end{equation}

\subsection{A construction of multi-field E model}
\noindent The $F$ term potential of E model with K\"ahler potential $K$ and the super-potential $W$ is given by
\begin{equation}
\frac{ {K}_{\text{E model}} }{ {M}_{\text{pl}}^{2} } = - \frac{1}{2} \sum_{l=1}^{n} 3 {\alpha}_{l} \left( \frac{ {T}_{l} - \overline{{T}_{l}} }{ {T}_{l} + \overline{{T}_{l}} } \right)^{2} + \frac{ \left| {W}_{0} \right|^{2} }{ \left| {F}_{S} \right|^{2} + h \left( {T}_{1} + \overline{{T}_{1}}, \cdots, {T}_{n} + \overline{{T}_{n}} \right) } S \overline{S}, \quad \quad W = {W}_{0} {e}^{S} - \left( \delta + \frac{1}{3} \right) {W}_{0}, \\
\end{equation}
\noindent where $h \left( 1 - \frac{{T}_{1} + \overline{{T}_{1}}}{2}, \cdots, 1 - \frac{{T}_{n} + \overline{{T}_{n}}}{2} \right)$ is a real function with $h \left( 0, \cdots, 0 \right) = 0$, $\nabla h  \left( 0, \cdots, 0 \right) = \bold{0}$ and $\left. \frac{d^2 h  \left( {y}_{1}, \cdots, {y}_{n} \right)}{d {y}_{s} d {y}_{t} } \right|_{0} = 0$ $\forall s,t \in \left\{ 1, \cdots, n \right\}$ with $s \neq t$, evaluated at ${T}_{1I} = \cdots = {T}_{nI} = {S}_{R} = {S}_{I} = 0$, becomes
\begin{equation}
{V}_{\text{E model}} = \Lambda + \frac{h \left( 1 - {T}_{1R}, \cdots, 1 - {T}_{nR} \right)}{ {M}_{\text{pl}}^{2} }. \\
\end{equation}
\noindent where $\Lambda = \delta \left( 2 - 3 \delta \right) \frac{\left| {W}_{0} \right|^{2}}{ {M}_{\text{pl}}^{2} }$. Particularly, when we take 
\begin{equation}
\label{A special function of E model}
h \left( 1 - \frac{{T}_{1} + \overline{{T}_{1}}}{2}, \cdots, 1 - \frac{{T}_{n} + \overline{{T}_{n}}}{2} \right) = \sum_{l=1}^{n} {m}_{l}^{2} \left( 1 - \frac{{T}_{l} + \overline{{T}_{l}} }{2} \right)^{2 {n}_{l} } = \sum_{l=1}^{n} {m}_{l}^{2} \left( 1 - {e}^{- \sqrt{\frac{2}{3 {\alpha}_{l}}} \frac{ {\phi}_{l} }{ {M}_{\text{pl}} } } \right)^{2 {n}_{l}}, \\
\end{equation}
\noindent and in terms of canonical fields ${\phi}_{l}$, where $\text{Re}{\left( {T}_{l} \right)} = {e}^{- \sqrt{\frac{2}{3 {\alpha}_{l}}} \frac{ {\phi}_{l} }{ {M}_{\text{pl}} } }$, and take ${T}_{lI}=0$, it gives the multi-field E model 
\begin{equation}
\label{A Multi-field E model}
{V}_{\text{E model}} = \Lambda + {M}_{\text{pl}}^{-2} \sum_{l=1}^{n} {m}_{l}^{2} \left( 1 - {e}^{- \sqrt{\frac{2}{3 {\alpha}_{l}}} \frac{ {\phi}_{l} }{ {M}_{\text{pl}} } } \right)^{2 {n}_{l}}. \\
\end{equation}
\noindent Next, we are going to study their properties at their corresponding minimum point(s).

\section{Properties of minimum point(s)}
\label{Properties of minimum point(s)}
\subsection{T model}
\noindent Since the derivatives of the $T$ model potential Eq.(\ref{A Multi-field T model}) are
\begin{equation}
\frac{d {V}_{\text{T model}} }{d {\phi}_{l} } = \frac{2 {m}_{l}^{2} {n}_{l} } { {M}_{\text{pl}}^{3} } \sqrt{\frac{2}{3 {\alpha}_{l} } } \frac{ \tanh^{2 {n}_{l} }{\left( \frac{ {\phi}_{l} }{ {M}_{\text{pl}} \sqrt{6 {\alpha}_{l} } } \right)} }{ \sinh{\left( \sqrt{\frac{2}{3 {\alpha}_{l} }} \frac{ {\phi}_{l} }{ {M}_{\text{pl}} } \right)} }, \\
\end{equation}

\begin{equation}
\begin{split}
\frac{ {d}^{2} {V}_{\text{T model}} }{d {\phi}_{l}^{2} } =&\; - \frac{4 {m}_{l}^{2} {n}_{l} }{3 {\alpha}_{l} {M}_{\text{pl}}^{4} } \left[ - 2 {n}_{l} + \cosh{\left( \sqrt{\frac{2}{3 {\alpha}_{l} }} \frac{{\phi}_{l}}{ {M}_{\text{pl}} } \right)} \right] \frac{ \tanh^{2 {n}_{l} }{\left( \frac{ {\phi}_{l} }{ {M}_{\text{pl}} \sqrt{6 {\alpha}_{l} } } \right)} }{ \sinh^{2}{\left( \sqrt{\frac{2}{3 {\alpha}_{l} }} \frac{{\phi}_{l}}{{M}_{\text{pl}}} \right)} }, \\
\frac{ {d}^{2} {V}_{\text{T model}} }{d {\phi}_{i} d {\phi}_{j} } =&\; \frac{ {d}^{2} {V}_{\text{T model}} }{ d {\phi}_{j} d {\phi}_{i} } = 0, \quad \quad \quad \forall \; i, j \in \left\{1, \cdots, n \right\}, \; i \neq j, \\
\end{split}
\end{equation}
\noindent the field coordinates of the minimum are $\left( {\phi}_{1}, \cdots, {\phi}_{n} \right) = \left( 0,\cdots, 0 \right)$ (the origin). The elements of Hessian matrix evaluated at the minimum point are
\begin{equation}
\frac{d^2 {V}_{\text{T model}} }{d {\phi}_{l}^{2}} = 
\begin{cases}
 \frac{ {m}_{l}^{2} }{ 3 {\alpha}_{l} {M}_{\text{pl}}^{4} }, &\quad \quad \text{if} \quad {n}_{l} = 1; \\
 0, &\quad \quad \text{if} \quad {n}_{l} > 1, \\
 \end{cases}
\end{equation}
\noindent Since the kinetic terms are canonical, the elements of mass matrix evaluated at the minimum point are equal to that of Hessian matrix evaluated at the minimum point. The SUSY breaking scales evaluated at the vacuum are
\begin{equation}
{M}_{{Z}_{l}}^{4} := \left. {e}^{\frac{K}{ {M}_{\text{pl}}^{2} }} \left( {K}^{ {Z}_{l} \overline{{Z}_{l}} } {D}_{{Z}_{l}} W \overline{ {D}_{{Z}_{l}} W } \right) \right|_{0} = 0, \\
\end{equation}

\begin{equation}
{M}_{S}^{4} := \left. {e}^{\frac{K}{{M}_{\text{pl}}^{2}}} \left( {K}^{ S \overline{S} } {D}_{S} W \overline{ {D}_{S} W } \right) \right|_{0} = \left. \frac{ \left\{ \left| {F}_{S} \right|^{2} + f \left( \left| {Z}_{1} \right|^{2}, \cdots, \left| {Z}_{n} \right|^{2} \right) \right\} }{ {M}_{\text{pl}}^{2} } \right|_{0} = \frac{ \left| {F}_{S} \right|^{2} }{ {M}_{\text{pl}}^{2} } = \frac{2}{3} \left( 2 - 3 \delta \right) \frac{\left| {W}_{0} \right|^{2}}{{M}_{\text{pl}}^{2}}, \\
\end{equation}
\noindent where the last equality holds when we consider Eq.(\ref{A special function of T model}), while the gravitino mass evaluated at the minimum is
\begin{equation}
{M}_{3/2}^{4} := \left. {e}^{\frac{K}{{M}_{\text{pl}}^{2}}} \frac{\left| {W} \right|^{2}}{{M}_{\text{pl}}^{2}} \right|_{0} = \frac{\left| {W}_{0} + k \right|^{2}}{{M}_{\text{pl}}^{2}} = \left( \frac{2}{3} - \delta \right)^{2} \frac{\left| {W}_{0} \right|^{2}}{{M}_{\text{pl}}^{2}}, \\
\end{equation}
\noindent resulting in $\sum_{l=1}^{n} {M}_{ {Z}_{l} }^{4} + {M}_{S}^{4} - 3 {M}_{3/2}^{4} = {M}_{\text{pl}}^{-2} \left\{ \left| {F}_{S} \right|^{2} - 3 \left| {W}_{0} + k \right|^{2} \right\} = \left. {V}_{\text{T model}} \right|_{0}$. Obviously, SUSY of ${Z}_{l}$ is unbroken while that of $S$ is broken with the scale $\left| {F}_{S} \right|^{1/2}$ at the minimum.  This result is independent of the number of constrained super-fields because the function $f \left( \left| {Z}_{1} \right|^{2}, \cdots, \left| {Z}_{n} \right|^{2} \right)$ is taken as a polynomial of ${Z}_{1}, \cdots, {Z}_{n}$ and it vanishes at the minimum point.

\subsection{E model}
\noindent Since the derivatives of the $E$ model potential Eq.(\ref{A Multi-field E model}) are
\begin{equation}
\frac{d {V}_{\text{E model}} }{d {\phi}_{l} } = \frac{ 2 {m}_{l}^{2} {n}_{l} }{ {M}_{\text{pl}}^{3} } \sqrt{\frac{2 }{3 {\alpha}_{l}} } {e}^{- \sqrt{\frac{2}{3 {\alpha}_{l} }} \frac{{\phi}_{l}}{ {M}_{\text{pl}} } } \left( 1 - {e}^{- \sqrt{\frac{2}{3 {\alpha}_{l} }} \frac{{\phi}_{l}}{{M}_{\text{pl}}} } \right)^{2 {n}_{l} - 1 }, \\
\end{equation}

\begin{equation}
\begin{split}
\frac{d^{2} {V}_{\text{E model}} }{d {\phi}_{l}^{2} } =& - \frac{ 4 {m}_{l}^{2} {n}_{l} }{ 3 {\alpha}_{l} {M}_{\text{pl}}^{4} } {e}^{- \sqrt{\frac{2}{3 {\alpha}_{l} }} \frac{{\phi}_{l}}{{M}_{\text{pl}}} } \left( 1 - 2 {n}_{l} {e}^{- \sqrt{\frac{2}{3 {\alpha}_{l} }} \frac{{\phi}_{l}}{{M}_{\text{pl}}} } \right) \left( 1 - {e}^{- \sqrt{\frac{2}{3 {\alpha}_{l} }} \frac{{\phi}_{l}}{{M}_{\text{pl}}} } \right)^{2 {n}_{l} - 2 }, \\
\frac{d^{2} {V}_{\text{E model}} }{d {\phi}_{i} d {\phi}_{j} } =& \frac{d^{2} {V}_{\text{E model}} }{d {\phi}_{j} d {\phi}_{i} } = 0, \quad \quad \quad \forall \; i, j \in \left\{1, \cdots, n \right\}, \; i \neq j, \\
\end{split}
\end{equation}
\noindent the field coordinates of the minimum are $\left( {\phi}_{1}, \cdots, {\phi}_{n} \right) = \left( 0,\cdots, 0 \right)$ (the origin). The elements of Hessian matrix evaluated at the minimum point are
\begin{equation}
\left. \frac{d^{2} {V}_{\text{E model}} }{d {\phi}_{l}^{2} } \right|_{0} =
\begin{cases}
 \frac{ 4 {m}_{l}^{2} }{ 3 {\alpha}_{l} {M}_{\text{pl}}^{4} }, &\quad \quad \text{if} \quad {n}_{l} = 1; \\
 0, &\quad \quad \text{if} \quad {n}_{l} > 1, \\
 \end{cases}
\end{equation}
\noindent Since the kinetic terms are canonical, the elements of mass matrix evaluated at the minimum point are equal to that of Hessian matrix evaluated at the minimum point. The SUSY breaking scales evaluated at the vacuum are
\begin{equation}
{M}_{{T}_{l}}^{4} := \left. {e}^{\frac{K}{{M}_{\text{pl}}^{2}}} \left( {K}^{ {T}_{l} \overline{{T}_{l}} } {D}_{{T}_{l}} W \overline{ {D}_{{T}_{l}} W } \right) \right|_{0} = 0, \\
\end{equation}

\begin{equation}
{M}_{S}^{4} := \left. {e}^{\frac{K}{{M}_{\text{pl}}^{2}}} \left( {K}^{ S \overline{S} } {D}_{S} W \overline{ {D}_{S} W } \right) \right|_{0} =\left. \frac{\left| {F}_{S} \right|^{2} + h \left(1 - {T}_{1}, \cdots, 1 - {T}_{n} \right) }{ {M}_{\text{pl}}^{2} } \right|_{0} = \frac{ \left| {F}_{S} \right|^{2}}{ {M}_{\text{pl}}^{2}} = \frac{2}{3} \left( 2 - 3 \delta \right) \frac{\left| {W}_{0} \right|^{2}}{{M}_{\text{pl}}^{2}}, \\
\end{equation}

\noindent where the equality holds when we consider Eq.(\ref{A special function of E model}), while the gravitino mass evaluated at the minimum is
\begin{equation}
{M}_{3/2}^{4} := \left. {e}^{\frac{K}{{M}_{\text{pl}}^{2}}} \frac{\left| {W} \right|^{2}}{{M}_{\text{pl}}^{2}} \right|_{0} = \frac{\left| {W}_{0} + k \right|^{2}}{{M}_{\text{pl}}^{2}} = \left( \frac{2}{3} - \delta \right)^{2} \frac{\left| {W}_{0} \right|^{2}}{{M}_{\text{pl}}^{2}}, \\
\end{equation}
\noindent resulting in $\sum_{l=1}^{n} {M}_{ {T}_{l} }^{4} + {M}_{S}^{4} - 3 {M}_{3/2}^{4} = {M}_{\text{pl}}^{-2} \left\{ \left| {F}_{S} \right|^{2} - 3 \left| {W}_{0} + k \right|^{2} \right\} = \left. {V}_{\text{E model}} \right|_{0}$. Obviously, SUSY of ${T}_{l}$ is unbroken while that of $S$ is broken with the scale $\left| {F}_{S} \right|^{1/2}$ at the minimum.  This result is independent of the number of constrained super-fields because the function $h \left( 1 - \frac{ {T}_{1} + \overline{{T}_{1}}}{2}, \cdots, 1 - \frac{ {T}_{n} + \overline{{T}_{n}}}{2} \right)$ is taken as a polynomial of $1 - \frac{ {T}_{1} + \overline{{T}_{1}}}{2}, \cdots, 1 - \frac{ {T}_{n} + \overline{{T}_{n}}}{2}$ and it vanishes at the minimum point. Given that we know their potential forms and properties at minimum, it is time to study the field evolutions to see how they describe inflation. Before that, let us recall the essential equations of motion for the subsequent evolution analysis.

\section{The basic setup of multi constrained fields}
\label{The basic setup of multi constrained fields}
\subsection{T model}
\noindent The bosonic part of Lagrangian of multi-field T model is
\begin{equation}
\label{Lagrangian of multi-field T model}
\begin{split}
\frac{\mathcal{L}_{\text{T model}} }{\sqrt{-g}} =&\; \frac{ {M}_{\text{pl}}^{2} }{2} R - {M}_{\text{pl}}^{2} \sum_{l=1}^{n} \frac{ 3 {\alpha}_{l} }{ \left( 1 -  {Z}_{lR}^{2} \right)^{2} } {\partial}_{\mu} {Z}_{lR} {\partial}^{\mu} {Z}_{lR} - \left[ {\Lambda} + {M}_{\text{pl}}^{-2} \sum_{l=1}^{n} {m}_{l}^{2} \left( {Z}_{lR}^{2} \right)^{2 {n}_{l} } \right] \\
=&\; \frac{ {M}_{\text{pl}}^{2} }{2} R - \frac{1}{2} \sum_{l=1}^{n} {\partial}_{\mu} {\phi}_{l} {\partial}^{\mu} {\phi}_{l} - \left[ {\Lambda} + {M}_{\text{pl}}^{-2} \sum_{l=1}^{n} {m}_{l}^{2} \tanh^{2 {n}_{l} }{\left( \frac{ {\phi}_{l} }{ {M}_{\text{pl}} \sqrt{6 {\alpha}_{l} } } \right)} \right], \\
\end{split}
\end{equation}

\noindent where ${\Lambda} = {M}_{\text{pl}}^{-2} \left[ \left| {F}_{S} \right|^{2} - 3 \left| {W}_{0} + k \right|^{2} \right]$ and the second equality holds after the substitution of canonically normalized fields. We expand the fields to the first order around its classical background values 
\begin{equation}
\label{Decomposition of background fields and perturbations}
{\phi}_{l} \left( {x}^{\mu} \right) = {\phi}_{lb} \left( t \right) + {\delta} {\phi}_{lb} \left( {x}^{\mu} \right). \\
\end{equation}
\noindent The norm of the velocity vector is given by the background components of the fields
\begin{equation}
\label{The norm of the velocity vector}
\dot{\sigma}^{2} = \sum_{l=1}^{n} \dot{\phi}_{lb}^{2} \quad \Rightarrow \quad \dot{\sigma} = \sqrt{ \sum_{l=1}^{n} \dot{\phi}_{lb}^{2} }, \\
\end{equation}
\noindent where the norm $\dot{\sigma}$ is defined to be positive. The background components of fields depend on the cosmic time $t$ only (or the number of e-foldings $N$ defined as $dN = H dt$) and we can have the Laplacian as
\begin{equation}
\square {\phi}_{lb} = - \left( \ddot{\phi}_{lb} + 3 H \dot{\phi}_{lb} \right) = - H^2 \left[ \frac{d^2 {\phi}_{lb} }{ d N^2 } + \left( 3 - {\epsilon}_{H} \right) \frac{d {\phi}_{lb} }{ d N } \right], \\
\end{equation}
\noindent where ${\epsilon}_{H} = - \frac{\dot{H}}{H^2}$ is the first order Hubble slow-roll parameter, and the equations of motion (E.O.M.)s of the background components ${\phi}_{lb}$ are\footnote{For numerical simulation, the E.O.M.s become
\begin{equation}
V \frac{d^2 {\phi}_{l} }{d N^2} + \frac{V}{ {M}_{\text{pl}}^{2} } \left( 3 {M}_{\text{pl}}^{2} - \frac{1}{2} {\sigma'}^{2} \right) \frac{d {\phi}_{l} }{d N} + \left( 3 {M}_{\text{pl}}^{2} - \frac{1}{2} {\sigma'}^{2} \right) \frac{d V}{d {\phi}_{l} } = 0. \\
\end{equation}}
\begin{equation}
\begin{split}
&\;\ddot{\phi}_{lb} +3 H \dot{ {\phi}_{lb} } + \sqrt{ \frac{8}{3 {\alpha}_{l} } } \frac{ {m}_{l}^{2} {n}_{l} }{ {M}_{\text{pl}}^{3} } \frac{ \tanh^{2 {n}_{l}}{ \left( \frac{ {\phi}_{lb} }{ {M}_{\text{pl}} \sqrt{6 {\alpha}_{l} } } \right) } }{ \sinh{\left( \sqrt{\frac{2}{3 {\alpha}_{l} }} \frac{{\phi}_{lb}}{{M}_{\text{pl}}} \right)} } \\
=&\; H^2 \left[ \frac{d^2 {\phi}_{lb} }{ d N^2 } + \left( 3 - {\epsilon}_{H} \right) \frac{d {\phi}_{lb} }{ d N } \right] + \sqrt{ \frac{8}{3 {\alpha}_{l} } } \frac{{m}_{l}^{2} {n}_{l} }{{M}_{\text{pl}}^{3}}\frac{ \tanh^{2 {n}_{l}}{ \left( \frac{ {\phi}_{lb} }{ {M}_{\text{pl}} \sqrt{6 {\alpha}_{l} } } \right) } }{ \sinh{\left( \sqrt{\frac{2}{3 {\alpha}_{l} }} \frac{{\phi}_{lb}}{{M}_{\text{pl}}} \right)} } = 0. \\
\end{split}
\end{equation}

\subsection{E model}
\noindent The bosonic part of Lagrangian of multi-field E model is
\begin{equation}
\label{Lagrangian of multi-field E model}
\begin{split}
\frac{\mathcal{L}_{\text{E model}} }{\sqrt{-g}} =&\; \frac{ {M}_{\text{pl}}^{2} }{2} R - {M}_{\text{pl}}^{2} \sum_{l=1}^{n} \frac{ 3 {\alpha}_{l} }{ 4 {T}_{lR}^{2} } {\partial}_{\mu} {T}_{lR} {\partial}^{\mu} {T}_{lR} - \left[ {\Lambda} + {M}_{\text{pl}}^{-2} \sum_{l=1}^{n} {m}_{l}^{2} \left( 1- {T}_{lR} \right)^{2 {n}_{l} } \right] \\
=&\; \frac{{M}_{\text{pl}}^{2}}{2} R - \frac{1}{2} \sum_{l=1}^{n} {\partial}_{\mu} {\phi}_{l} {\partial}^{\mu} {\phi}_{l} - \left[ {\Lambda} + {M}_{\text{pl}}^{-2} \sum_{l=1}^{n} {m}_{l}^{2} \left( 1- {e}^{- \sqrt{\frac{2}{3 {\alpha}_{l} }} {\phi}_{l} } \right)^{2 {n}_{l} } \right], \\
\end{split}
\end{equation}

\noindent After the decomposition of the fields into classical background values and perturbations, the norm of the velocity vector is the same as Eq.(\ref{The norm of the velocity vector}), and the E.O.M.s becomes
\begin{equation}
\begin{split}
&\; \ddot{\phi}_{lb} +3 H \dot{ {\phi}_{lb} } + \frac{2 {m}_{l}^{2} {n}_{l}}{{M}_{\text{pl}}^{3}} \sqrt{\frac{2}{3 {\alpha}_{l} }} {e}^{- \sqrt{\frac{2}{3 {\alpha}_{l} }} \frac{{\phi}_{l}}{{M}_{\text{pl}}} } \left( 1 - {e}^{- \sqrt{ \frac{2}{3 {\alpha}_{l} } } \frac{{\phi}_{l}}{{M}_{\text{pl}}} } \right)^{2 {n}_{l} - 1} \\
=&\; H^2 \left[ \frac{d^2 {\phi}_{lb} }{ d N^2 } + \left( 3 - {\epsilon}_{H} \right) \frac{d {\phi}_{lb} }{ d N } \right] + \frac{2 {m}_{l}^{2} {n}_{l}}{ {M}_{\text{pl}}^{3} } \sqrt{\frac{2}{3 {\alpha}_{l} }} {e}^{- \sqrt{\frac{2}{3 {\alpha}_{l} }} \frac{{\phi}_{l}}{{M}_{\text{pl}}} } \left( 1 - {e}^{- \sqrt{ \frac{2}{3 {\alpha}_{l} } } \frac{{\phi}_{l}}{{M}_{\text{pl}}} } \right)^{2 {n}_{l} - 1} = 0. \\
\end{split}
\end{equation}
\noindent Next, we are going to evaluate the double field inflation dynamics of T and E models respectively.

\section{Numerical calculations}
\label{Numerical calculations}
\noindent In this section, we take double field models as examples to show the trajectory, the scale of turn rate and the square mass of the entropic perturbation. Since the inflation dynamics is irrelevant to $\left| {W}_{0} \right|^{2}$ in these cases, we do not give a numerical number to it. 

\subsection{T model}
\begin{table}[h!]
\begin{center}
\begin{tabular}{ |c|c|c|c|c|c|c|c|c|c|c|c|c|c|c|c| }
\hline
$\Lambda/ {M}_{\text{pl}}^{4}$ & ${\alpha}_{1}$ & ${\alpha}_{2}$ & ${m}_{1}/ {M}_{\text{pl}}^{3}$ & ${m}_{2}/ {M}_{\text{pl}}^{3}$ & ${n}_{1}$ & ${n}_{2}$ & ${\phi}_{1\text{ini}}/{M}_{\text{pl}}$ \\
\hline
$0$ & $2$ & $2$ & $1.1 \times 10^{-5}$ & $1 \times 10^{-5}$ & $1$ & $1$ & $6$ \\
\hline
${\phi}_{2\text{ini}}/{M}_{\text{pl}}$ & ${\phi'}_{1\text{ini}}/{M}_{\text{pl}}$ & ${\phi'}_{2\text{ini}}/{M}_{\text{pl}}$ & ${N}_{\text{end}}$ & ${N}_{\text{stop}}$ & ${\beta}_{\text{iso}}$ & $\cos{\Delta}$ & $$ \\
\hline
$4.8$ & $1.5 \times 10^{-3}$ & $1.5 \times 10^{-3}$ & $50.8923$ & $51.6790$ & $7.11473 \times {10}^{-36}$ & $0.120559$ & $$ \\
\hline
\end{tabular}
\end{center}
\caption{Parameters and initial conditions for a double field T model. ${N}_{\text{end}}$ is the number of e-foldings when ${\epsilon}_{H} = 1$, while ${N}_{\text{stop}}$ is the number of e-foldings that we stop for numerical calculation. ${\beta}_{\text{iso}}$ and $\cos{\Delta}$ are defined in Eq.(\ref{betaiso}) and Eq.(\ref{Trigo Delta}) respectively. Since the cosmological constant $\Lambda$ is small ($\Lambda \approx {10}^{-120} {M}_{\text{pl}}^{4}$), we can phenomenologically take $\Lambda = 0$. }
\label{table: Parameters and initial conditions for a double field T model}
\end{table}

\begin{figure}[h!]
\centering
\includegraphics[width=80mm, height=60mm]{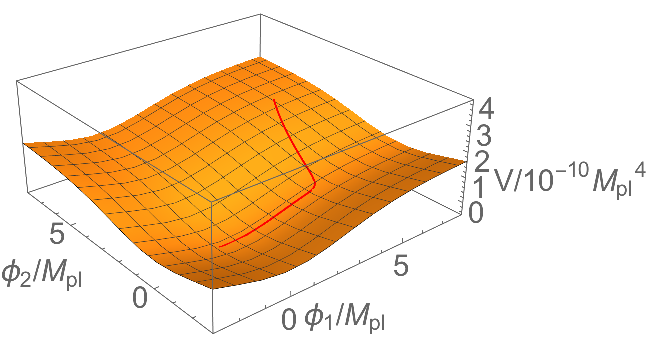} 
\includegraphics[width=80mm, height=60mm]{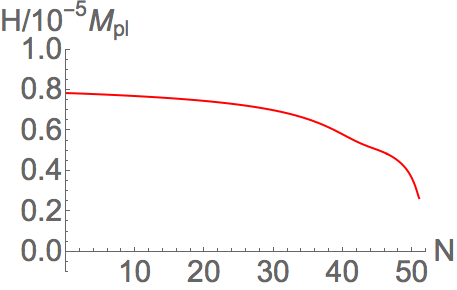} 
\caption{Left: The trajectory and the potential of T model. Right: Hubble parameter evolution of T model. The parameters are listed in Table \ref{table: Parameters and initial conditions for a double field T model}. } 
\label{fig: T2 trajectory and potential}
\end{figure}

\begin{figure}[h!]
\centering
\includegraphics[width=80mm, height=60mm]{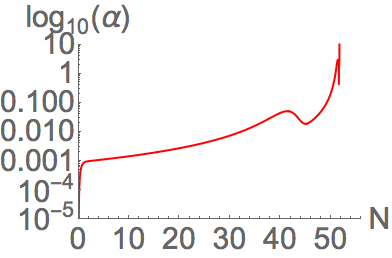} 
\includegraphics[width=80mm, height=60mm]{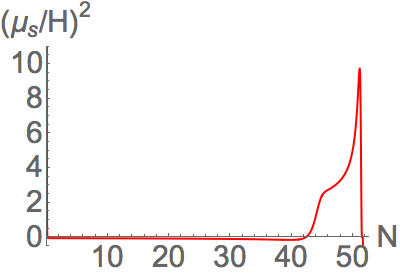} 
\caption{Left: $2$ times Turn rate per Hubble $\alpha = 2 \frac{\omega}{H} $ of T model. Right: Square mass of the entropic perturbation $\left( \frac{ {\mu}_{s} }{H} \right)^{2}$ of T model. The parameters are listed in Table \ref{table: Parameters and initial conditions for a double field T model}. }
\label{fig: T2 Turn rate}
\end{figure}

\noindent The possible parameter set of T model satisfying the e-folding constraint $50 \leq {N}_{\text{end}} - {N}_{\text{hc}} \leq 60$ is listed in Table \ref{table: Parameters and initial conditions for a double field T model}. From Figure \ref{fig: T2 trajectory and potential}, starting from a slope, the trajectory rolls down a nearly straight line to a valley along ${\phi}_{1}$ direction, turns significantly to move along the valley and then reaches the minimum point for oscillation. This significant turning is shown as a bump in the $\log_{10}{\left( \alpha \right)} - N$ graph in Figure \ref{fig: T2 Turn rate} at about $42$ e-foldings since the speed of field evolution ${\sigma}'$ drops at that point. 

\vspace{3mm}

\noindent Apart from this, the square mass of the entropic perturbation $\left( {\mu}_{s}/ H \right)^{2}$, which is given by Eq.(\ref{square mass of entropic perturbation}), is shown in Figure \ref{fig: T2 Turn rate}. One can see that before $42$ e-foldings, $\left( {\mu}_{s}/ H \right)^{2}$ remains small and negative. Since ${\mu}_{s}^{2}$ is related to the field curvature of the potential in the direction orthogonal to the trajectory, light and negative values mean that the field curvature orthogonal to the trajectory is light and negative so that the trajectory is rolling. At $42$ e-foldings, $\left( {\mu}_{s}/ H \right)^{2}$ surges since the trajectory turns to a valley, which means the field curvature becomes relatively large and positive. And finally, the trajectory runs to the minimum point for oscillation.

\subsection{E model}
\begin{table}[h!]
\begin{center}
\begin{tabular}{ |c|c|c|c|c|c|c|c|c|c|c|c|c|c|c|c| }
\hline
$\Lambda/{M}_{\text{pl}}^{4}$ & ${\alpha}_{1}$ & ${\alpha}_{2}$ & ${m}_{1}/{M}_{\text{pl}}^{3}$ & ${m}_{2}/{M}_{\text{pl}}^{3}$ & ${n}_{1}$ & ${n}_{2}$ & ${\phi}_{1\text{ini}}/{M}_{\text{pl}}$ \\
\hline
$0$ & $2$ & $2$ & $1.1 \times 10^{-5}$ & $1 \times 10^{-5}$ & $1$ & $1$ & $5.8$ \\
\hline
${\phi}_{2\text{ini}}/{M}_{\text{pl}}$ & ${\phi'}_{1\text{ini}}/{M}_{\text{pl}}$ & ${\phi'}_{2\text{ini}}/{M}_{\text{pl}}$ & ${N}_{\text{end}}$ & ${N}_{\text{stop}}$ & ${\beta}_{\text{iso}}$ & $\cos{\Delta}$ & $$ \\
\hline
$4.7$ & $2 \times 10^{-3}$ & $2 \times 10^{-3}$ & $54.7826$ & $56.9525$ & $2.96101 \times {10}^{-39}$ & $0.168902$ & $$ \\
\hline
\end{tabular}
\end{center}
\caption{Parameters and initial conditions for a double field E model. ${N}_{\text{end}}$ is the number of e-foldings when ${\epsilon}_{H} = 1$, while ${N}_{\text{stop}}$ is the number of e-foldings that we stop for numerical calculation. ${\beta}_{\text{iso}}$ and $\cos{\Delta}$ are defined in Eq.(\ref{betaiso}) and Eq.(\ref{Trigo Delta}) respectively. Since the cosmological constant $\Lambda$ is small ($\Lambda \approx {10}^{-120} {M}_{\text{pl}}^{4}$), we can phenomenologically take $\Lambda = 0$. }
\label{table:Parameters and initial conditions for a double field E model}
\end{table}

\begin{figure}[h!]
\centering
\includegraphics[width=100mm, height=50mm]{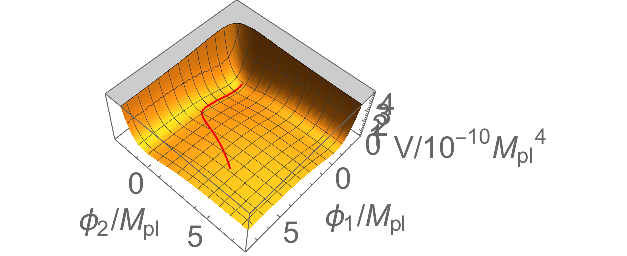} 
\includegraphics[width=60mm, height=50mm]{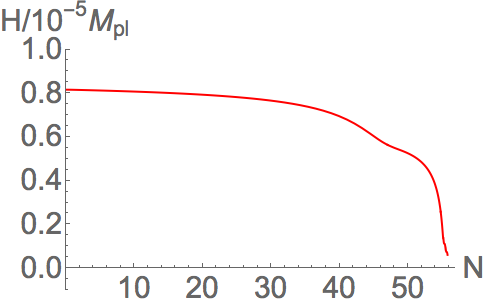} 
\caption{Left: The trajectory and the potential of E model. Right: Hubble parameter evolution of E model. The parameters are listed in Table \ref{table:Parameters and initial conditions for a double field E model}. } 
\label{fig: E2 trajectory and potential}
\end{figure}

\begin{figure}[h!]
\centering
\includegraphics[width=80mm, height=60mm]{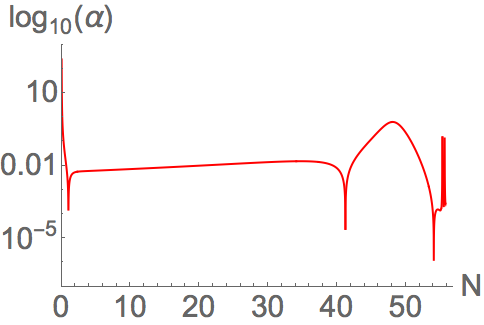} 
\includegraphics[width=80mm, height=60mm]{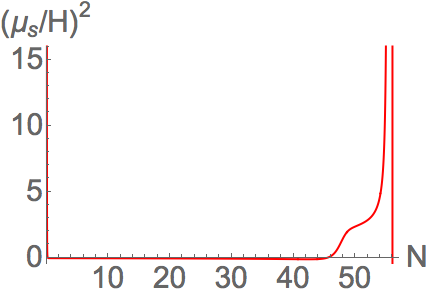} 
\caption{Left: $2$ times turn rate per Hubble $\alpha = 2 \frac{\omega}{H} $ of E model (in log scale). Right: Square mass of the entropic perturbation $\left( \frac{ {\mu}_{s} }{H} \right)^{2}$ of E model. The parameters are listed in Table \ref{table:Parameters and initial conditions for a double field E model}. }
\label{fig: E2 Turn rate}
\end{figure}

\noindent The possible parameter set of E model satisfying the e-folding constraint $50 \leq {N}_{\text{end}} - {N}_{\text{hc}} \leq 60$ is listed in Table \ref{table:Parameters and initial conditions for a double field E model}. From Figure \ref{fig: E2 trajectory and potential}, starting from a slope, the trajectory rolls down a nearly straight line to a valley along ${\phi}_{1}$ direction, turns significantly to move along the valley and then reaches the minimum point for oscillation. This significant turning is shown as a bump in the $\log_{10}{\left( \alpha \right)} - N$ graph in Figure \ref{fig: E2 Turn rate} at about $48$ e-foldings since the the speed of field evolution ${\sigma}'$ drops at that point. 

\vspace{3mm}

\noindent Apart from this, the square mass of the entropic perturbation $\left( {\mu}_{s}/ H \right)^{2}$, which is given by Eq.(\ref{square mass of entropic perturbation}), is shown in Figure \ref{fig: E2 Turn rate}. One can see that before $48$ e-foldings, $\left( {\mu}_{s}/ H \right)^{2}$ remains small and negative. Since ${\mu}_{s}^{2}$ is related to the field curvature of the potential in the direction orthogonal to the trajectory, light and negative values mean that the field curvature orthogonal to the trajectory is light and negative so that the trajectory is rolling. At $42$ e-foldings, $\left( {\mu}_{s}/ H \right)^{2}$ surges since the trajectory turns to a valley, which means the field curvature becomes relatively large and positive. And finally, the trajectory runs to the minimum point for oscillation.

\section{Discussion}
\label{Discussion}
\subsection{Similarity of T/E models and path dependence on initial conditions}
\begin{figure}[h!]
\centering
\includegraphics[width=90mm, height=60mm]{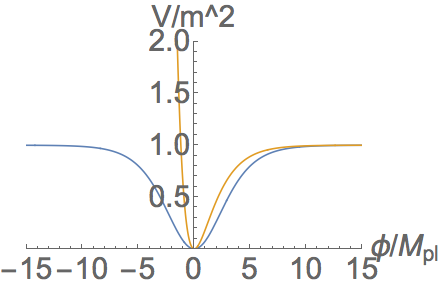} 
\caption{The $1$D potentials of $T$ and $E$ models (normalized by $m^2$). ($T$ model (blue): ${m}^2 \tanh^{2 p}{\left( \frac{\phi }{ {M}_{\text{pl}} \sqrt{6 {\alpha} }} \right)}$, $E$ model (yellow): ${m}^2 \left( 1 - {e}^{- \sqrt{\frac{2}{3 \alpha} } \frac{\phi}{{M}_{\text{pl}}} } \right)^{2 p}$. In this figure, $\alpha = 2$, $p=1$. )} 
\label{fig: Potential_Compare}
\end{figure}

\noindent We take two fields for the investigation to show the patterns of turning. Since the shapes of $T$ and $E$ models are very similar in the range of inflation (about $0 \leq \phi \leq 6 {M}_{\text{pl}} $) as we can see in Figure \ref{fig: Potential_Compare}, and the corresponding kinetic terms are canonical, the physics of scalar fields in both models will be similar. In addition, the shape of the trajectory depends on the initial conditions. For instance, one may set the starting point with coordinates ${\phi}_{1} < {\phi}_{2}$. In that case, the trajectory slides along negative ${\phi}_{1}$ direction first and turns to the valley along ${\phi}_{2}$ direction instead to reach the minimum point.

\subsection{An advantage of considering orthogonal nilpotent super-fields}
\noindent In fact, these $T$ and $E$ models can be obtained by other motivations such as \cite{1803.09911} and \cite{1502.07733}. The difference between the approaches in \cite{1803.09911} and \cite{1901.09046} is that in the approach like \cite{1803.09911}, one needs to assume the mass scale of the associated part of the scalar field responsible for inflation is sufficiently heavy such that it is stabilized throughout the inflation process, while the approach of \cite{1901.09046} provides a physical meaning to the sinflaton that its mass scale is very large. This gives a more physical reason for us to ignore the associated part of the scalar field, which makes us easier to construct potentials with more physical meanings.

\subsection{Comments on the modified super-potential Eq.(\ref{modified super-potential})}
\noindent There are some comments pointing out that the $S$ components (real and imaginary parts) of the first derivatives evaluated at the minimum need not be zero because it is sufficient to consider the approximately zero of the first derivatives and $S$ can be written as a bilinear spinor form. Here our logic is the following. Since the origin of the orthogonal nilpotent constraints comes from the finite limits in the Lagrangian when the masses of sgoldstino, sinflaton and inflatino are very large, $S$ and $\text{Im} \left( \Phi \right)$ should be normally considered as scalar fields instead of the bilinear spinor forms before the E.O.M. calculation of super-fields. Since they are scalar fields, they should also be stabilized at the minimum. The minimum point of T model is $\left({Z}_{R}, {Z}_{I}, {S}_{R}, {S}_{I} \right) = \left(0,0,0,0 \right)$ while that of E model is $\left({T}_{R}, {T}_{I}, {S}_{R}, {S}_{I} \right) = \left(1,0,0,0 \right)$ and similar for multi-field versions. Thus, we consider whether there are some modifications to make the $S$ components of the first derivatives vanish, thereby showing the results above.

\section{Conclusions}
\label{Conclusions}
\noindent To conclude, our modification on the construction of T/E models can make all the components of the first derivative exactly zero at the minimum point. The modified part can be arbitrarily taken to obtain a physical cosmological constant. We also extend the modified T/E models into the multi-field versions by considering multi orthogonal nilpotent super-fields, whose sgoldstino, sinflatons and inflatinos are very heavy such that orthogonal nilpotent constraints are taken for the sake of obtaining physical limits. Finally, we study the double inflation dynamics of both T and E models respectively, and show their turning patterns, turning rate scales (about $O \left( 0.1 \right)$), the square mass of the entropic perturbation with scales $O \left( 1 \right)$ and relative transfer function $\cos{\Delta}$ with values $0.12$ and $0.169$ for T and E models respectively. These models can be further verified by updated observation data, and by considering other induced cosmological phenomena like the possibility of production of primordial black hole due to the trajectory turning \cite{2006.16641, 2005.02895, 2004.08369, 2004.06106}, which can be one of the directions of future work.

\section{Ackonwledgement}
\noindent The author thanks Prof. Hiroyuki ABE and Dr. Shuntaro Aoki very much for suggestion and useful discussions. The author is supported by Scholarship for Young Doctoral Students, Waseda University, Academic Year 2019.

\appendix
\section{A review: Single orthogonal constrained field}
\label{A review: Single orthogonal constrained field}
\subsection{A construction of T model}
\noindent Basically, $\bold{S}$ and $\bold{\Phi}$ satisfying the nilpotent and orthogonality condition as
\begin{equation}
\label{Constraints on super-fields of T model}
\bold{S}^{2} = 0, \quad \bold{S} \bold{B} = 0, \\
\end{equation}
\noindent where $\bold{B} = \frac{1}{2 i} \left( \bold{\Phi} - \overline{\bold{\Phi}} \right)$. The K\"ahler function $\mathcal{G}$ is given by
\begin{equation}
\mathcal{G} = - \frac{1}{2} \frac{3 \alpha}{\left( 1 - Z \overline{Z} \right)^{2}} \left( Z - \overline{Z} \right)^{2} + \frac{\left| {W}_{0} \right|^{2}}{\left| {F}_{S} \right|^{2} + f \left( Z \overline{Z} \right) } S \overline{S} + S + \overline{S} + \ln{ \left( \frac{\left| {W}_{0} \right|^{2}}{ {M}_{\text{pl}}^{6} } \right) }, \\
\end{equation}

\noindent or equivalently, the K\"ahler potential and super-potential are\footnote{Note that under the constraints, we obtain
\begin{equation}
- \frac{3 {\alpha} }{2} \ln{\left[ \frac{\left( 1 - Z \overline{Z} \right)^{2} }{\left( 1 - {Z}^{2} \right) \left( 1 - \overline{Z}^{2} \right)} \right]} = - \frac{3 {\alpha}}{2} \left[ \frac{ \left( Z - \overline{Z} \right)^{2}}{ \left( 1 - Z \overline{Z} \right)^{2} } \right]. \\
\end{equation}}
\begin{equation}
\label{Kahler potential of T model}
\frac{K}{{M}_{\text{pl}}^{2}} = - \frac{3 {\alpha} }{2} \ln{\left[ \frac{\left( 1 - Z \overline{Z} \right)^{2} }{\left( 1 - {Z}^{2} \right) \left( 1 - \overline{Z}^{2} \right)} \right]} + \frac{\left| {W}_{0} \right|^{2}}{\left| {F}_{S} \right|^{2} + f \left( Z \overline{Z} \right) } S \overline{S}, \quad \quad W = {W}_{0} {e}^{S}, \\
\end{equation}

\noindent The $F$ term potential is
\begin{equation}
{V}_{F} := {M}_{\text{pl}}^{4} {e}^{\mathcal{G}} \left( \mathcal{G}^{{\alpha} \overline{\beta}} \mathcal{G}_{\alpha} \mathcal{G}_{\overline{\beta}} - 3 \right) = {M}_{\text{pl}}^{-2} \left[ \left| {F}_{S} \right|^{2} - 3 \left| {W}_{0} \right|^{2} + f \left( {Z}^{2} \right) \right]. \\
\end{equation}

\noindent When we take\footnote{One may check that $f\left( 0 \right) = 0$. } $f \left( Z \overline{Z} \right) = m^2 \left( Z \overline{Z} \right)^{n}$, in terms of a canonical field $\phi$, where $\text{Re}\left( Z \right) = \tanh{ \left( \frac{\phi}{ {M}_{\text{pl}} \sqrt{6 {\alpha}}} \right)}$, we have
\begin{equation}
\label{Single field T model}
{V}_{F} := \frac{ \left| {F}_{S} \right|^{2} - 3 \left| {W}_{0} \right|^{2} }{ {M}_{\text{pl}}^{2} } + \frac{m^2}{ {M}_{\text{pl}}^{2} } \tanh^{2n}{\left( \frac{\phi}{ {M}_{\text{pl}} \sqrt{6 {\alpha}}} \right)}, \\
\end{equation}
\noindent leading to the $T$ model potential given in \cite{1807.06211}. In particular, the simple $T$ model described in \cite{1901.09046} can be obtained by taking $n=1$. 

\subsection{A construction of E model}
\noindent We impose the constraints
\begin{equation}
\label{Constraints on super-fields of E model}
\bold{S}^{2} = 0, \quad \bold{S} \left( \bold{T} - \overline{\bold{T}} \right) = 0, \quad \Rightarrow \quad \left( \bold{T} - \overline{\bold{T}} \right)^{3} = 0, \\
\end{equation}
\noindent on the super-fields of the $E$ model. The K\"ahler function $\mathcal{G}$ is
\begin{equation}
\mathcal{G} = - \frac{1}{2} \frac{3 {\alpha} }{\left( T + \overline{T} \right)^{2} } \left( T - \overline{T} \right)^{2} + \frac{ \left| {W}_{0} \right|^{2} }{ \left| {F}_{S} \right|^{2} + h \left( 1 - \frac{T + \overline{T} }{2} \right) } S \overline{S} + S + \overline{S} + \ln{\left( \frac{\left| {W}_{0} \right|^{2}}{{M}_{\text{pl}}^{6}} \right) }, \\
\end{equation}
\noindent or equivalently, the K\"ahler potential and super-potential are\footnote{Note that under the constraints, we obtain
\begin{equation}
- \frac{3 {\alpha} }{2} \ln{\left[ \frac{ \left( T + \overline{T} \right)^{2} }{ 4 T \overline{T} } \right]} = - \frac{3 {\alpha} }{2} \left( \frac{ T - \overline{T} }{ T + \overline{T} } \right)^{2}. \\
\end{equation}}
\begin{equation}
\label{Kahler potential of E model}
\frac{K}{{M}_{\text{pl}}^{2}} =  - \frac{3 {\alpha} }{2} \ln{\left[ \frac{ \left( T + \overline{T} \right)^{2} }{ 4 T \overline{T} } \right]} + \frac{ \left| {W}_{0} \right|^{2} }{ \left| {F}_{S} \right|^{2} + h \left( 1 - \frac{T + \overline{T} }{2} \right) } S \overline{S}, \quad \quad W = {W}_{0} {e}^{S}. \\
\end{equation}

\noindent This $T$ parametrized K\"ahler potential can be obtained by taking the following Cayley transformation
\begin{equation}
T = \frac{Z+1}{-Z+1} \quad \Leftrightarrow \quad Z = \frac{T-1}{T+1}, \\
\end{equation}
\noindent from the $Z$ parametrized K\"ahler potential and vice versa without using the super-field constraints Eq.(\ref{Constraints on super-fields of T model}) and Eq.(\ref{Constraints on super-fields of E model}) since
\begin{equation}
- \frac{3 {\alpha} }{2} \ln{\left[ \frac{\left( 1 - Z \overline{Z} \right)^{2} }{\left( 1 - {Z}^{2} \right) \left( 1 - \overline{Z}^{2} \right)} \right]} = - \frac{3 {\alpha} }{2} \ln{\left[ \frac{ \left( T + \overline{T} \right)^{2} }{ 4 T \overline{T} } \right]}. \\
\end{equation}

\noindent The $F$ term potential is
\begin{equation}
{V}_{F} = \frac{ \left| {F}_{S} \right|^{2} - 3 \left| {W}_{0} \right|^{2} }{ {M}_{\text{pl}}^{2} } + \frac{1}{ {M}_{\text{pl}}^{2} } h \left( 1 - \frac{T + \overline{T} }{2} \right), \\
\end{equation}

\noindent or terms of a canonical field $\phi$, where $\text{Re} \left( T \right) = {e}^{- \sqrt{\frac{2}{3 \alpha}} \frac{\phi}{{M}_{\text{pl}}} }$,
\begin{equation}
{V}_{F} = \frac{ \left| {F}_{S} \right|^{2} - 3 \left| {W}_{0} \right|^{2} }{ {M}_{\text{pl}}^{2} } + \frac{1}{ {M}_{\text{pl}}^{2} } h \left( 1 - {e}^{- \sqrt{\frac{2}{3 \alpha}} \frac{\phi}{{M}_{\text{pl}}} } \right), \\
\end{equation}

\noindent By taking\footnote{One may check that $h \left( 0 \right) = h' \left( 0 \right) = 0$. }
\begin{equation}
h \left( 1 - \frac{T + \overline{T} }{2} \right) = m^2 \left( 1 - \frac{T + \overline{T} }{2} \right)^{2 n}, \\
\end{equation}
\noindent we obtain the $E$ model potential given in \cite{1807.06211}
\begin{equation}
{V}_{F} = \frac{ \left| {F}_{S} \right|^{2} - 3 \left| {W}_{0} \right|^{2} }{ {M}_{\text{pl}}^{2} } + \frac{m^2}{ {M}_{\text{pl}}^{2} } \left( 1 - \frac{T + \overline{T} }{2} \right)^{2 n} = \frac{ \left| {F}_{S} \right|^{2} - 3 \left| {W}_{0} \right|^{2} }{ {M}_{\text{pl}}^{2} } + \frac{m^2}{ {M}_{\text{pl}}^{2} } \left( 1 - {e}^{- \sqrt{\frac{2}{3 \alpha}} \frac{\phi}{{M}_{\text{pl}}} } \right)^{2 n}. \\
\end{equation} 
\noindent In particular, the simple $E$ model described in \cite{1901.09046} can be obtained by taking $n=1$.

\section{A derivation of the modified constant}
\label{A derivation of the modified constant}
\noindent We keep the original K\"ahler potential and modify the super-potential by adding a constant $k \in \mathbb{C}$ \cite{2001.09574} as
\begin{equation}
\label{modified super-potential}
W = {W}_{0} e^{S} + k. \\
\end{equation}
\noindent In this case, the $F$ term potential of T model at $S = {Z}_{I} = 0$ becomes
\begin{equation}
{V}_{\text{T model}} = \frac{ \left| {F}_{S} \right|^{2} - 3 \left| {W}_{0} + k \right|^{2} }{ {M}_{\text{pl}}^{2} } + \frac{ f \left( {Z}_{R}^{2} \right)}{ {M}_{\text{pl}}^{2} }, \\
\end{equation}
\noindent and its first derivative evaluated at the minimum point becomes
\begin{equation}
{M}_{\text{pl}}^{-2} \{0,0,2 [ \left| {F}_{S} \right|^{2} - 2 \left| {W}_{0} \right|^{2} - 2 \text{Re} \left( \overline{{W}_{0}} k \right) + f(0) ], - 4 \text{Im} \left( \overline{{W}_{0}} k \right) \}. \\
\end{equation}

\noindent We can see that
\begin{equation}
\label{A cosmological constant condition}
\begin{split}
\text{Im} \left( \overline{{W}_{0}} k \right) =&\; 0, \\
\left| {F}_{S} \right|^{2} - 3 \left| {W}_{0} + k \right|^{2} =&\; {M}_{\text{pl}}^{2} \Lambda, \\
\left| {F}_{S} \right|^{2} - 2 \left| {W}_{0} \right|^{2} - 2 \text{Re} \left( \overline{{W}_{0}} k \right) =&\; 0. \\
\end{split}
\end{equation}
\noindent On solving, we obtain 
\begin{equation}
\begin{split}
\Lambda =&\; - \frac{\overline{W}_{0} }{ {W}_{0} } \frac{\left( {W}_{0} + k \right) \left( {W}_{0} + 3 k \right)}{ {M}_{\text{pl}}^{2} }, \\
\left| {F}_{S} \right|^{2} =&\; 2 \left( k + {W}_{0} \right) \overline{{W}_{0}}. \\
\end{split}
\end{equation}
\noindent If\footnote{This trivially satisfies $\text{Im} \left( \overline{{W}_{0}} k \right) = 0$. } $ - {W}_{0} < k \leq - \frac{1}{3} {W}_{0}$, we can attain the dS spacetime, and accordingly, all the first derivatives can be zero provided that $\left| {F}_{S} \right|^{2} = 2 \left( k + {W}_{0} \right) \overline{{W}_{0}}$. For example, when we take $k = - \frac{2}{3} {W}_{0}$, we obtain $\Lambda = \frac{1}{3} \frac{ \left| {W}_{0} \right|^{2} }{ {M}_{\text{pl}}^{2} }$ and $\left| {F}_{S} \right|^{2} = \frac{2}{3} \left| {W}_{0} \right|^{2}$. Hence, this can be applicable when $\Lambda$, $\left| {W}_{0} \right|^{2}$ and $\left| {F}_{S} \right|^{2}$ have the same scale. Interestingly, when we take $k$ sufficiently close to $- \frac{1}{3} {W}_{0}$, say, $k = - \left( \delta + \frac{1}{3} \right) {W}_{0}$ with sufficiently small $\delta > 0$, we obtain $\Lambda = \delta \left( 2 - 3 \delta \right) \frac{ \left| {W}_{0} \right|^{2} }{ {M}_{\text{pl}}^{2} }$ and $\left| {F}_{S} \right|^{2} = \frac{2}{3} \left( 2 - 3 \delta \right) \left| {W}_{0} \right|^{2}$. The smallness of $\delta$ can be responsible for that of $\Lambda$, and allow arbitrary scales for $\left| {W}_{0} \right|^{2}$ and $\left| {F}_{S} \right|^{2}$. This implies when $\delta$ is sufficiently small, it can allow $\left| {W}_{0} \right|^{2}$ and $\left| {F}_{S} \right|^{2}$ much larger than $\Lambda$, while $\delta$ can have a scale $O\left(0.1 \right)$ with $\delta<\frac{2}{3}$ when $\left| {W}_{0} \right|^{2}$ and $\left| {F}_{S} \right|^{2}$ are compatible with $\Lambda$. Thus, without loss of generality, we take $k = - (\delta + \frac{1}{3}) {W}_{0}$ with $\delta > 0$ for discussion. We can also solve the problem in E model by this trick. The $F$ term potential of E model at $S = {T}_{I} = 0$ becomes
\begin{equation}
{V}_{\text{E model}} = \frac{ \left| {F}_{S} \right|^{2} - 3 \left| {W}_{0} + k \right|^{2} }{ {M}_{\text{pl}}^{2} } + \frac{ h \left( 1 - {T}_{R} \right)}{ {M}_{\text{pl}}^{2} }, \\
\end{equation}

\noindent and its first derivative evaluated at the minimum point becomes
\begin{equation}
{M}_{\text{pl}}^{-2} \left\{0,0,2 \left[ \left| {F}_{S} \right|^{2} - 2 \left| {W}_{0} \right|^{2} - 2 \text{Re} \left( \overline{{W}_{0}} k \right) + h(0) \right], - 4 \text{Im} \left( \overline{{W}_{0}} k \right) \right\}. \\
\end{equation}
\noindent By repeating the same procedures above, we can obtain the same result. Next, we extend these two models into multi-field cases.

\section{A formalism of Double Field Inflation}
\label{A formalism of Double Field Inflation}
\noindent In this section, we follow the derivation in \cite{1210.7487} and \cite{1310.8285}. For a recent application, please refer to \cite{1810.10546}. The action in Jordan frame is
\begin{equation}
\label{Action in Jordan frame}
{S}_{\text{Jordan}} = \int d^{4} x \sqrt{-\tilde{g}} \left[ f\left({\phi}^{I}\right) \tilde{R} - \frac{1}{2} \tilde{\mathcal{G}}_{IJ} \tilde{g}^{{\mu}{\nu}} \tilde{\triangledown}_{\mu} {\phi}^{I} \tilde{\triangledown}_{\nu} {\phi}^{J} - \tilde{V}\left( {\phi}^{I} \right) \right]. \\
\end{equation}
\noindent where $f\left( {\phi}^{I} \right)$ is the non-minimal coupling function and $\tilde{V}\left( {\phi}^{I} \right)$ is the potential for the scalar fields in Jordan frame. To change the action in Jordan frame into the counterpart in Einstein frame, we define a spacetime metric in Einstein frame ${g}_{{\mu}{\nu}}\left( x \right)$ as 
\begin{equation}
{g}_{{\mu}{\nu}}\left( x \right) = {\Omega}^{2} \left( x \right) \tilde{g}_{{\mu}{\nu}}\left( x \right), \\
\end{equation}
\noindent where the conformal factor ${\Omega}^{2} \left( x \right)$ is given by
\begin{equation}
{\Omega}^{2} \left( x \right) = \frac{2}{{M}^{2}_{\text{pl}}} f \left( {\phi}^{I} \left( x \right) \right). \\
\end{equation}
\noindent Then, the action in Jordan frame becomes that in Einstein frame, which is given by
\begin{equation}
\label{Action in Einstein frame}
{S}_{\text{Einstein}} = \int d^{4} x \sqrt{-{g}} \left[ \frac{{M}^{2}_{\text{pl}}}{2} {R} - \frac{1}{2} {\mathcal{G}}_{IJ} {g}^{{\mu}{\nu}} {\triangledown}_{\mu} {\phi}^{I} {\triangledown}_{\nu} {\phi}^{J} - {V}\left( {\phi}^{I} \right) \right]. \\
\end{equation}
\noindent The potential in Einstein frame becomes
\begin{equation}
\label{Potential in Einstein frame}
V \left( {\phi}^{I} \right) = \frac{\tilde{V} \left( {\phi}^{I} \right)}{{\Omega}^{4} \left( x \right)} = \frac{{M}^{4}_{\text{pl}}}{4 f^2 \left( {\phi}^{I} \right)} \tilde{V} \left( {\phi}^{I} \right). \\
\end{equation}
\noindent The coefficients $\mathcal{G}_{{I}{J}}$ of the non-canonical kinetic terms in Einstein frame depend on the non-minimal coupling function $f\left( {\phi}^{I} \right)$ and its derivatives. They are given by
\begin{equation}
\label{Field space metric transformation}
\mathcal{G}_{{I}{J}} \left( {\phi}^{K} \right) = \frac{{M}^{2}_{\text{pl}}}{2 f \left( {\phi}^{L} \right)} \left[ \tilde{\mathcal{G}}_{{I}{J}} \left( {\phi}^{K} \right) + \frac{3}{f \left( {\phi}^{L} \right)} {f}_{,I} {f}_{,J} \right], \\
\end{equation}
\noindent where ${f}_{,I}=\frac{{\partial}{f}}{{\partial}{\phi}^{I}}$. Varying the action in Einstein frame with respect to ${g}_{{\mu}{\nu}} \left( x \right)$, we have Einstein equations
\begin{equation}
{R}_{{\mu}{\nu}} - \frac{1}{2}{g}_{{\mu}{\nu}} R = \frac{1}{{M}^{2}_{\text{pl}}} {T}_{{\mu}{\nu}}, \\
\end{equation}
\noindent where 
\begin{equation}
{T}_{{\mu}{\nu}} = \mathcal{G}_{{I}{J}} {\partial}_{\mu} {\phi}^{I} {\partial}_{\nu} {\phi}^{J} - {g}_{{\mu}{\nu}} \left[ \frac{1}{2} \mathcal{G}_{{K}{L}} {\partial}_{\gamma} {\phi}^{K} {\partial}^{\gamma} {\phi}^{L} + V\left( {\phi}^{K} \right) \right]. \\
\end{equation}
\noindent Varying Eq. (\ref{Action in Einstein frame}) with respect to ${\phi}^{I}$, we obtain the equation of motion for ${\phi}^{I}$
\begin{equation}
\square {\phi}^{I} + {g}^{{\mu}{\nu}} {\Gamma}^{I}_{{J}{K}} {\partial}_{\mu} {\phi}^{J} {\partial}_{\nu} {\phi}^{K} - \mathcal{G}^{{I}{K}} {V}_{,K} = 0, \\ 
\end{equation}

\noindent where $\square {\phi}^{I} = {g}^{{\mu}{\nu}} {\phi}^{I}_{;{\mu}{\nu}}$ and ${\Gamma}^{I}_{{J}{K}}$ is the Christoffel symbol for the field space manifold in terms of $\mathcal{G}_{{I}{J}}$ and its derivatives. We expand each scalar field to the first order around its classical background value, 
\begin{equation}
{\phi}^{I} \left( {x}^{\mu} \right) = {\varphi}^{I} \left( t \right) + {\delta} {\phi}^{I} \left( {x}^{\mu} \right), \\
\end{equation}
\noindent and perturb a spatially flat Friedmann-Robertson-Walker (FRW) metric,
\begin{equation}
ds^2={g}_{{\mu}{\nu}} d{x}^{\mu} d{x}^{\nu} = - \left(1+2A \right) dt^2 + 2 a \left( {\partial}_{i} B \right) d{x}^{i} dt + a^2 \left[ \left( 1-2{\psi} \right) {\delta}_{ij} + 2 {\partial}_{i} {\partial}_{j} E \right] d{x}^{i} d{x}^{j}, \\
\end{equation}
\noindent where $a\left( t \right)$ is the scale factor. To the zeroth order, the $00$ and $ij$ components of Einstein equations become
\begin{equation}
H^2 = \frac{1}{3 {M}^{2}_{\text{pl}}} \left[ \frac{1}{2} \mathcal{G}_{{I}{J}} \dot{\varphi}^{I} \dot{\varphi}^{J} + V \left( {\varphi}^{I} \right) \right], \\
\end{equation}
\begin{equation}
\dot{H} = - \frac{1}{2 {M}^{2}_{\text{pl}}} \mathcal{G}_{{I}{J}} \dot{\varphi}^{I} \dot{\varphi}^{J}, \\
\end{equation}
\noindent where $H = \frac{\dot{a} \left( t \right)}{a \left( t \right)}$ is the Hubble parameter, and the field field space metric is calculated at the zeroth order, $\mathcal{G}_{{I}{J}} = \mathcal{G}_{{I}{J}} \left( {\varphi}^{K} \right)$. In terms of the number of e-foldings\footnote{In some literatures like \cite{1310.8285}, ${N}_{*} = {N}_{\text{tot}} - N\left( t \right)$ is used and the corresponding differential equation becomes $d {N}_{*} = - H dt$. But, in this paper, we keep using $dN=Hdt$.} $N=\ln{a}$ with $d N = H dt$, the above Einstein equation becomes
\begin{equation}
3 {M}^{2}_{\text{pl}} - \frac{1}{2} \mathcal{G}_{{I}{J}} {{\varphi}^{I}}' {{\varphi}^{J}}' = \frac{V \left( {\varphi}^{I} \right)}{H^2}, \\
\end{equation}
\begin{equation}
\frac{H'}{H} = - \frac{1}{2 {M}^{2}_{\text{pl}}} \mathcal{G}_{{I}{J}} {{\varphi}^{I}}' {{\varphi}^{J}}', \\
\end{equation}
\noindent where the prime $'$ means the derivative with respect to $N$. For any vector in the field space $A^{I}$, we define a covariant derivative with respect to the field-space metric as usual by
\begin{equation}
\mathcal{D}_{J} {A}^{I} = {\partial}_{J} {A}^{I} + {\Gamma}^{I}_{{J}{K}} {A}^{K}, \\
\end{equation}
\noindent and the time derivative with respect to the cosmic time $t$ is given by
\begin{equation}
\mathcal{D}_{t} {A}^{I} \equiv \dot{\varphi}^{J} \mathcal{D}_{J} {A}^{I} = \dot{A}^{I} + {\Gamma}^{I}_{{J}{K}} \dot{\varphi}^{J} {A}^{K} = H \left( {{A}^{I}}' + {\Gamma}^{I}_{{J}{K}} {{\varphi}^{J}}' {A}^{K} \right). \\
\end{equation}
\noindent Now, we define the length of the velocity vector for the background fields as
\begin{equation}
|\dot{\varphi}^{I}| \equiv \dot{\sigma} = \sqrt{\mathcal{G}_{{P}{Q}} \dot{\varphi}^{P} \dot{\varphi}^{Q}} \quad \Rightarrow \quad |{{\varphi}^{I}}'| \equiv {\sigma}' = \sqrt{\mathcal{G}_{{P}{Q}} {{\varphi}^{P}}' {{\varphi}^{Q}}'}. 
\end{equation}
\noindent After introducing the unit vector of the velocity vector of the background fields
\begin{equation}
\hat{\sigma}^{I} \equiv \frac{\dot{\varphi}^{I}}{\dot{\sigma}} = \frac{{{\varphi}^{I}}'}{{\sigma}'} = \frac{{{\varphi}^{I}}'}{\sqrt{\mathcal{G}_{{P}{Q}} {{\varphi}^{P}}' {{\varphi}^{Q}}'}}, \\
\end{equation}
\noindent the $00$ and $ij$ components of Einstein equations become
\begin{equation}
H^2 = \frac{1}{3 {M}^{2}_{\text{pl}}} \left[ \frac{1}{2} \dot{\sigma}^2 + V \right] \quad \Leftrightarrow \quad 3 {M}^{2}_{\text{pl}} - \frac{1}{2} {{\sigma}'}^{2} = \frac{V \left( {\varphi}^{I} \right)}{H^2} \quad \Leftrightarrow \quad \frac{V}{{M}_{\text{pl}}^{2} H^2} = \left( 3 - {\epsilon}_{H} \right), \\
\end{equation}
\begin{equation}
\dot{H} = - \frac{1}{2 {M}^{2}_{\text{pl}}} \dot{\sigma}^{2} \quad \Leftrightarrow \quad \frac{H'}{H} = - \frac{1}{2 {M}^{2}_{\text{pl}}} {{\sigma}'}^{2} \quad \Leftrightarrow \quad \frac{ {{\sigma}'}^{2} }{ {M}_{\text{pl}}^{2} } = \frac{\dot{\sigma}^{2}}{{M}_{\text{pl}}^{2} H^2} = 2 {\epsilon}_{H}, \\
\end{equation}
\noindent and the equation of motion of ${\phi}^{I}$ in the zeroth order is 
\begin{equation}
\ddot{\sigma} + 3 H \dot{\sigma} + {V}_{,\sigma} = 0 \quad \Leftrightarrow \quad \frac{\ddot{\sigma}}{H \dot{\sigma}} = - 3 - \frac{3 - {\epsilon}_{H} }{2 {\epsilon}_{H} } \frac{d}{d N} \left( \ln{V} \right), \\
\end{equation}
\noindent where
\begin{equation}
{V}_{,\sigma} \equiv \hat{\sigma}^{I} {V}_{,I}. \\
\end{equation}
\noindent and ${\epsilon}_{H}$ is the first order Hubble slow-roll parameter defined in Eq.(\ref{1st order Hubble slow-roll parameter}). We define a quantity $\hat{s}^{{I}{J}}$
\begin{equation}
\hat{s}^{{I}{J}} \equiv \mathcal{G}^{{I}{J}} - \hat{\sigma}^{I} \hat{\sigma}^{J}, \\
\end{equation}
\noindent which obeys the following relations with $\hat{\sigma}^{I}$
\begin{equation}
\begin{split}
\hat{\sigma}_{I} \hat{\sigma}^{I} = 1, \\
\hat{s}^{{I}{J}} \hat{s}_{{I}{J}} = \mathcal{N} - 1, \\
\hat{s}^{I}_{\; A} \hat{s}^{A}_{\; J} = \hat{s}^{I}_{\; J}, \\
\hat{\sigma}_{I} \hat{s}^{{I}{J}} = 0 \quad \forall J. \\
\end{split}
\end{equation}
\noindent The slow-roll parameters are given by
\begin{equation}
\label{1st order Hubble slow-roll parameter}
{\epsilon}_{H} \equiv - \frac{\dot{H}}{H^2} = \frac{3 \dot{\sigma}^{2} }{\dot{\sigma}^2 + 2 V} \quad \Leftrightarrow \quad \frac{ \dot{\sigma}^{2} }{V} = \frac{2 {\epsilon}_{H} }{3 - {\epsilon}_{H} }, \\
\end{equation}
\noindent and

\begin{equation}
{\eta}_{{\sigma}{\sigma}} \equiv {M}^{2}_{\text{pl}} \frac{\mathcal{M}_{{\sigma}{\sigma}}}{V} \quad \text{and} \quad {\eta}_{{s}{s}} \equiv {M}^{2}_{\text{pl}} \frac{\mathcal{M}_{{s}{s}}}{V}, \\
\end{equation}
\noindent where $\mathcal{M}^{I}_{J}$ is the effective mass squared matrix given by
\begin{equation}
\label{Effective mass-squared matrix}
\begin{split} 
{\mathcal{M}}^{I}_{J} \equiv&\; \mathcal{G}^{IK} \left( \mathcal{D}_{J} \mathcal{D}_{K} V \right) - \mathcal{R}^{I}_{LMJ} \dot{\varphi}^{L} \dot{\varphi}^{M}, \\
{\mathcal{M}}_{{\sigma}{J}} \equiv&\; \hat{\sigma}_{I} \mathcal{M}^{I}_{J} = \hat{\sigma}^{K} \left( \mathcal{D}_{K} \mathcal{D}_{J} V \right), \\
{\mathcal{M}}_{{\sigma}{\sigma}} \equiv&\; \hat{\sigma}_{I} \hat{\sigma}^{J} \mathcal{M}^{I}_{J} = \hat{\sigma}^{K} \hat{\sigma}^{J} \left( \mathcal{D}_{K} \mathcal{D}_{J} V \right), \\
{\mathcal{M}}_{{s}{J}} \equiv&\; \hat{s}_{I} \mathcal{M}^{I}_{J} = \hat{s}_{I} \left( \mathcal{G}^{IK} \left( \mathcal{D}_{J} \mathcal{D}_{K} V \right) - \mathcal{R}^{I}_{LMJ} \dot{\varphi}^{L} \dot{\varphi}^{M} \right), \\
{\mathcal{M}}_{{s}{s}} \equiv& \; \hat{s}_{I} \hat{s}^{J} \mathcal{M}^{I}_{J} = \hat{s}_{I} \hat{s}^{J} \left( \mathcal{G}^{IK} \left( \mathcal{D}_{J} \mathcal{D}_{K} V \right) - \mathcal{R}^{I}_{LMJ} \dot{\varphi}^{L} \dot{\varphi}^{M} \right), \\
\end{split}
\end{equation}
\noindent and $\hat{s}^{I}$ is defined in Eq.(\ref{s}). We define the turn-rate vector ${\omega}^{I}$ as the covariant rate of change of the unit vector $\hat{\sigma}^{I}$ and its square norm
\begin{equation}
{\omega}^{I} \equiv \mathcal{D}_{t} \hat{\sigma}^{I} = - \frac{1}{\dot{\sigma}} {V}_{,K} \hat{s}^{{I}{K}} = \frac{-1}{H {\sigma}'} {V}_{,K} \hat{s}^{{I}{K}}, \quad \quad {\omega}^2 := {\omega}_{L} {\omega}^{L} = \frac{1}{\dot{\sigma}^{2}} \left[ {V}_{,L} {V}^{,L} - {V}_{,{\sigma} }^{2} \right]. \\
\end{equation}
\noindent Since ${\omega}^{I} \propto \hat{s}^{{I}{K}}$, we have
\begin{equation}
{\omega}^{I} \hat{\sigma}_{I} = 0. \\
\end{equation}
\noindent We can also find 
\begin{equation}
\mathcal{D}_{t} \hat{s}^{{I}{J}} = - \hat{\sigma}^{I} {\omega}^{J} - \hat{\sigma}^{J} {\omega}^{I}. \\
\end{equation}
\noindent Also, we introduce a new unit vector $\hat{s}^{I}$ pointing in the direction of the turn-rate, ${\omega}^{I}$, and a new projection operator ${\gamma}^{{I}{J}}$
\begin{equation}
\label{s}
\hat{s}^{I} \equiv \frac{{\omega}^{I}}{\omega}, \\
\end{equation}
\begin{equation}
{\gamma}^{{I}{J}} \equiv {\mathcal{G}}^{{I}{J}} - \hat{\sigma}^{I} \hat{\sigma}^{J} - \hat{s}^{I} \hat{s}^{J}. \\
\end{equation}
\noindent where ${\omega} = |{\omega}^{I}|$ is the magnitude of the turn-rate vector. The new unit vector $\hat{s}^{I}$ and the new projection operator ${\gamma}^{{I}{J}}$ also satisfy
\begin{equation}
\begin{split}
\hat{s}^{{I}{J}} =& \hat{s}^{I} \hat{s}^{J} + {\gamma}^{{I}{J}}, \\
{\gamma}^{{I}{J}} {\gamma}_{{I}{J}} =& \mathcal{N} -2, \\
\hat{s}^{{I}{J}} \hat{s}_{J} =& \hat{s}^{I}, \\
\hat{\sigma}_{I} \hat{s}^{I} =& \hat{\sigma}_{I} {\gamma}^{{I}{J}} = \hat{s}_{I} {\gamma}^{{I}{J}} = 0 \quad \forall J.
\end{split}
\end{equation}
\noindent We then find 
\begin{equation}
{\mathcal{D}}_{t} \hat{s}^{I} = - {\omega} \hat{\sigma}^{I} - {\Pi}^{I}, \\
{\mathcal{D}}_{t} {\gamma}^{{I}{J}} = \hat{s}^{I} {\gamma}^{J} + \hat{s}^{J} {\gamma}^{I}, \\
\end{equation}
\noindent where
\begin{equation}
{\Pi}^{I} \equiv \frac{1}{\omega} {\mathcal{M}}_{{\sigma}{K}} {\gamma}^{{I}{K}}, \\
\end{equation}
\noindent and hence
\begin{equation}
\hat{\sigma}_{I} {\Pi}^{I} = \hat{s}_{I} {\Pi}^{I} = 0. \\
\end{equation}
\noindent Now, we define the curvature and entropic perturbations as follows
\begin{equation}
\label{Definition of curvature perturbation}
\mathcal{R} = {\psi} + \frac{H}{\dot{\sigma}} \hat{\sigma}_{J} {\delta} {\phi}^{J} = \frac{H}{\dot{\sigma}} {Q}_{\sigma}, \\
\end{equation}
\begin{equation}
\label{Definition of entropic perturbation}
\mathcal{S} = \frac{H}{\dot{\sigma}} {Q}_{s}, \\
\end{equation}

\noindent whose E.O.M.s are given by \cite{1310.8285}
\begin{equation}
\begin{split}
\ddot{Q}_{\sigma} + 3 H \dot{Q}_{\sigma} + \left[ \left( \frac{k}{a} \right)^{2} + \mathcal{M}_{{\sigma}{\sigma}} - {\omega}^{2} - \frac{1}{{M}_{\text{pl}}^{2} a^3} \frac{d}{dt} \left( \frac{a^3 \dot{\sigma}^{2}}{H} \right) \right] {Q}_{\sigma} =&\; 2 \frac{d}{dt} \left( {\omega} {Q}_{s} \right) - 2 \left( \frac{{V}_{, \sigma}}{\dot{\sigma}} + \frac{\dot{H}}{H} \right) {\omega} {Q}_{s}, \\
\ddot{Q}_{s} + 3 H \dot{Q}_{s} + \left[ \left( \frac{k}{a} \right)^{2} + \mathcal{M}_{ss} + 3 {\omega}^{2} \right] {Q}_{s} =&\; 4 {M}_{\text{pl}}^{2} \frac{\omega}{\dot{\sigma}} \frac{k^2}{a^2} {\Psi}, \\
\end{split}
\end{equation}
\noindent where $\Psi$ is the gauge-invariant Bardeen potential \cite{astro-ph/0507632, 0809.4944}, $\mathcal{M}_{{\sigma}{\sigma}}$ and $\mathcal{M}_{ss}$ are given by Eq.(\ref{Effective mass-squared matrix}) and

\begin{equation}
\label{Effective mass squared of entropic perturbation}
{\mu}_{s}^{2} = \mathcal{M}_{ss} + 3 {\omega}^{2}, \\
\end{equation}
\noindent is the (effective) square mass of entropic perturbations. After the first horizon crossing, the co-moving wave number $k$ obeys $\frac{k}{aH}<1$. Hence, the curvature and entropic perturbations satisfy the following equations
\begin{equation}
\label{An evolution equation of curvature perturbation}
\dot{\mathcal{R}} = {\alpha} H \mathcal{S} + O \left( \frac{k^2}{a^2 H^2} \right), \\
\end{equation}
\begin{equation}
\label{An evolution equation of entropic perturbation}
\dot{\mathcal{S}} = {\beta} H \mathcal{S} + O \left( \frac{k^2}{a^2 H^2} \right), \\
\end{equation}
\noindent which allow us to write the transfer functions
\begin{equation}
\label{TRS integration in terms of cosmic time}
{T}_{\mathcal{RS}} \left( {t}_{\text{hc}}, t \right) = \int^{t}_{{t}_{\text{hc}}} d {t}' {\alpha} \left( t' \right) H \left( t' \right) {T}_{\mathcal{SS}} \left( {t}_{\text{hc}}, t' \right), \\
\end{equation}
\begin{equation}
\label{TSS integration in terms of cosmic time}
{T}_{\mathcal{SS}} \left( {t}_{\text{hc}}, t \right) = \exp \left[ \int^{t}_{{t}_{\text{hc}}} {dt'} {\beta} \left( t' \right) H \left( t' \right) \right], \\
\end{equation}
\noindent where ${t}_{\text{hc}}$ is the time of the first horizon crossing. Being changed from the cosmic time $t$ into the number of e-foldings $N$, ${T}_{\mathcal{RS}} \left( {t}_{\text{hc}}, t \right)$ and ${T}_{\mathcal{SS}} \left( {t}_{\text{hc}}, t \right)$ become 
\begin{equation}
\label{TRS integration in terms of e-foldings}
{T}_{\mathcal{RS}} \left( {N}_{\text{hc}}, N \right) = \int^{N}_{{N}_{\text{hc}}} {dN'} {\alpha} \left( N' \right) {T}_{\mathcal{SS}} \left( {N}_{\text{hc}}, N' \right), \\
\end{equation}
\noindent and
\begin{equation}
\label{TSS integration in terms of e-foldings}
{T}_{\mathcal{SS}} \left( {N}_{\text{hc}}, N \right) = \exp \left[ \int^{N}_{{N}_{\text{hc}}} {dN'} {\beta} \left( N' \right) \right]. \\
\end{equation}
\noindent The E.O.M.s of curvature and entropic perturbations are \cite{1210.7487}
\begin{equation}
\label{Differential equation of curvature perturbation}
\dot{\mathcal{R}} = 2 {\omega} \mathcal{S} + O \left( \frac{k^2}{a^2 H^2} \right), \\
\end{equation}
\noindent and
\begin{equation}
\label{Differential equation of entropic perturbation}
\dot{Q}_{s} \simeq - \frac{{\mu}^{2}_{s}}{3 H} {Q}_{s}, \\
\end{equation}
\noindent where the square mass of entropic perturbation can be written as 
\begin{equation}
\label{square mass of entropic perturbation}
{\mu}^{2}_{s} = {\mathcal{M}}_{ss} + 3 {\omega}^{2} \quad \Leftrightarrow \quad \frac{{\mu}^{2}_{s}}{H^2} = \left( 3 - {\epsilon} \right) {\eta}_{ss} + \frac{3}{4} {\alpha}^{2}, \\
\end{equation}
\noindent and $\simeq$ means slow-roll approximation and ${\alpha}$ is given in Eq.(\ref{alpha}). Comparing with Eq.(\ref{Definition of curvature perturbation}), (\ref{Definition of entropic perturbation}), (\ref{An evolution equation of curvature perturbation}) and (\ref{An evolution equation of entropic perturbation}) with Eq.(\ref{Differential equation of curvature perturbation}) and (\ref{Differential equation of entropic perturbation}) \cite{1210.7487}, we obtain
\begin{equation}
\label{alpha}
{\alpha} \left( t \right) = \frac{2 {\omega} \left( t \right)}{H \left( t \right)} \quad \Leftrightarrow \quad {\alpha} \left( N \right) = \frac{2 {\omega} \left( N \right)}{H \left( N \right)}, \\
\end{equation}
\noindent and 
\begin{equation}
{\beta} = - \frac{{\mu}^{2}_{s}}{3 H^2} - {\epsilon} - \frac{\ddot{\sigma}}{H \dot{\sigma}} = - {\eta}_{ss} \left( 1 - \frac{1}{3}{\epsilon} \right) + \left( 3 - {\epsilon} \right) + \frac{3 - {\epsilon}}{2 {\epsilon}} \frac{d}{d N} \left({\ln{V}} \right) - \frac{1}{4} {\alpha}^{2}, \\
\end{equation}
\noindent The power spectrum for the gauge invariant curvature perturbation is given by
\begin{equation}
\langle \mathcal{R} \left( \bold{k}_{1} \right) \mathcal{R} \left( \bold{k}_{2} \right) \rangle = \left( 2 \pi \right)^{3} {\delta}^{\left( 3 \right)} \left( \bold{k}_{1} + \bold{k}_{2} \right) {P}_{\mathcal{R}} \left( {k}_{1} \right), \\
\end{equation}
\noindent where ${P}_{\mathcal{R}} \left( k \right) = |\mathcal{R}|^{2}$. The dimensionless power spectrum is 
\begin{equation}
{\mathcal{P}}_{\mathcal{R}} = \frac{k^3}{2 {\pi}^2} |\mathcal{R}|^{2}, \\
\end{equation}
\noindent and the spectral index is defined as
\begin{equation}
{n}_{s} \equiv 1 + \left. \frac{d \ln {\mathcal{P}}_{\mathcal{R}}}{d \ln {k}} \right|_{\text{hc}}, \\
\end{equation}
\noindent where $k$ is the pivot scale\footnote{The pivot scale $k$ is related to the cosmic time $t$ by \begin{equation}
\frac{d \ln{k}}{dt} = \frac{d \left(aH \right)}{d t} = \frac{\dot{a}}{a} + \frac{\dot{H}}{H} = H \left( 1 + \frac{\dot{H}}{H^2} \right) = \left( 1 - {\epsilon}_{H} \right) H. \\
\end{equation}} and ${\text{hc}}$ means the first horizon crossing. Using the transfer function, we can relate the power spectra of adiabatic and entropic perturbations at time ${t}_{\text{hc}}$ to their values at some later time $t > {t}_{\text{hc}}$ with the corresponding pivot scale $k$ as
\begin{equation}
\begin{split}
{\mathcal{P}}_{\mathcal{R}} \left( k \right) =& \; {\mathcal{P}}_{\mathcal{R}} \left( {k}_{\text{hc}} \right) \left[ 1 + {T}^{2}_{\mathcal{RS}} \left( {t}_{\text{hc}}, t \right) \right], \\
{\mathcal{P}}_{\mathcal{S}} \left( k \right) =& \; {\mathcal{P}}_{\mathcal{R}} \left( {k}_{\text{hc}} \right) {T}^{2}_{\mathcal{SS}} \left( {t}_{\text{hc}}, t \right). \\
\end{split}
\end{equation}
\noindent The transfer functions satisfy 
\begin{equation}
\begin{split}
\frac{1}{H \left( {t}_{\text{hc}} \right)} \frac{{\partial} {T}_{\mathcal{RS}} \left( {t}_{\text{hc}}, t \right) }{{\partial} {t}_{\text{hc}}} &=\; - {\alpha} \left( {t}_{\text{hc}} \right) - {\beta} \left( {t}_{\text{hc}} \right) {T}_{\mathcal{SS}} \left( {t}_{\text{hc}}, {t} \right), \\
\frac{1}{H \left( {t}_{\text{hc}} \right)} \frac{{\partial} {T}_{\mathcal{SS}} \left( {t}_{\text{hc}}, t \right) }{{\partial} {t}_{\text{hc}}} &=\; - {\beta} \left( {t}_{\text{hc}} \right) {T}_{\mathcal{SS}} \left( {t}_{\text{hc}}, t \right). \\
\end{split}
\end{equation}
\noindent In term of the number of e-foldings $N$, the above differential equations become
\begin{equation}
\begin{split}
\frac{{\partial} {T}_{\mathcal{RS}} \left( {N}_{\text{hc}}, N \right) }{ {\partial} {N}_{\text{hc}} } &=\; - {\alpha} \left( {N}_{\text{hc}} \right) - {\beta} \left( {N}_{\text{hc}} \right) {T}_{\mathcal{SS}} \left( {N}_{\text{hc}}, N \right), \\
\frac{{\partial} {T}_{\mathcal{SS}} \left( {N}_{\text{hc}}, N \right) }{{\partial} {N}_{\text{hc}}} &=\; - {\beta} \left( {N}_{\text{hc}} \right) {T}_{\mathcal{SS}} \left( {N}_{\text{hc}}, N \right). \\
\end{split}
\end{equation}
\noindent The spectral index for the power spectrum of the adiabatic fluctuations becomes
\begin{equation}
{n}_{s} \simeq {n}_{s} \left( {t}_{\text{hc}} \right) + \frac{1}{H} \left( \frac{{\partial} {T}_{\mathcal{RS}}}{{\partial} {t}_{\text{hc}}} \right) \sin{2 \Delta}, \\
\end{equation}
\noindent where
\begin{equation}
{n}_{s} \left( {t}_{\text{hc}} \right) = 1 - 6 {\epsilon}_{H} \left( {t}_{\text{hc}} \right) + 2 {\eta}_{{\sigma}{\sigma}} \left( {t}_{\text{hc}} \right), \\
\end{equation}
\noindent and the trigonometric functions for ${T}_{\mathcal{RS}}$ are defined as
\begin{equation}
\label{Trigo Delta}
\sin{\Delta} \equiv \; \frac{1}{\sqrt{1 + T^{2}_{\mathcal{RS}}}}, \quad \cos{\Delta} \equiv \frac{{T}_{\mathcal{RS}}}{\sqrt{1 + {T}^{2}_{\mathcal{RS}}}}, \quad \tan{\Delta} \equiv \; \frac{1}{T_{\mathcal{RS}}}. \\
\end{equation}
\noindent The iso-curvature fraction is given by
\begin{equation}
\label{betaiso}
{\beta}_{\text{iso}} \equiv \frac{{\mathcal{P}}_{\mathcal{S}}}{{\mathcal{P}}_{\mathcal{R}} + {\mathcal{P}}_{\mathcal{S}}} = \frac{T^{2}_{\mathcal{SS}}}{1 + T^{2}_{\mathcal{SS}} + T^{2}_{\mathcal{RS}}}, \\
\end{equation}
\noindent which can be used for comparing the predictions with the recent observation data. Also, the tensor-to-scalar ratio is given by
\begin{equation}
r \simeq \frac{16 {\epsilon}}{1 + T^{2}_{\mathcal{RS}}}. \\
\end{equation}

\end{document}